\documentclass[fleqn,usenatbib]{mnras}

\usepackage[T1]{fontenc}

\usepackage{amssymb}

\DeclareRobustCommand{\VAN}[3]{#2}
\let\VANthebibliography\thebibliography
\def\thebibliography{\DeclareRobustCommand{\VAN}[3]{##3}\VANthebibliography}

\usepackage{graphicx}	
\usepackage{amsmath}	
\usepackage{hyperref}

\title[\textit{Planck} SZiFi catalogues]{The \textit{Planck} SZiFi catalogues: a new set of \textit{Planck} catalogues of Sunyaev-Zeldovich-detected galaxy clusters}

\author[\'{I}. Zubeldia et al.]{
\'{I}\~{n}igo Zubeldia$^{1,2}$\thanks{inigo.zubeldia@ast.cam.ac.uk}, Jean-Baptiste Melin$^3$, Jens Chluba$^4$ and Richard Battye$^4$
\\
$^{1}$Institute of Astronomy, University of Cambridge, Madingley Road, Cambridge CB3 0HA, United Kingdom\\
$^{2}$Kavli Institute for Cosmology, University of Cambridge, Madingley Road, Cambridge CB3 0HA, United Kingdom
\\
$^3$Universit{\'e} Paris-Saclay, CEA, D{\'e}partement de Physique des Particules, 91191, Gif-sur-Yvette, France
\\
$^4$Jodrell Bank Centre for Astrophysics, Department of Physics and Astronomy, University of Manchester, Manchester, M13 9PL,\\ United Kingdom
}

\date{Accepted XXX. Received YYY; in original form ZZZ}

\pubyear{2024}

\begin{document}
\label{firstpage}
\pagerange{\pageref{firstpage}--\pageref{lastpage}}
\maketitle

\begin{abstract}
We introduce the \textit{Planck} \texttt{SZiFi} catalogues, a new set of 10 catalogues of galaxy clusters detected through their thermal Sunyaev-Zeldovich (tSZ) signature. The catalogues are produced by applying the \texttt{SZiFi} cluster finder to the \textit{Planck} PR3 temperature data down to a signal-to-noise threshold of 5. They span three frequency channel combinations ($100$--$857$\,GHz, $100$--$545$\,GHz, and $100$--$353$\,GHz) and 7 of them are constructed by spectrally deprojecting the Cosmic Infrared Background (CIB). This approach allows us, for the first time in the context of cluster finding, to carefully assess the impact of the cluster-correlated CIB on the recovered cluster tSZ observables, which we find to be negligible. In addition, we quantify the impact of the relativistic corrections to the tSZ signal, finding them to be at the 5--10\,\% level for the cluster tSZ amplitude but negligible for the signal-to-noise. We compile our catalogues into a single \textit{Planck} \texttt{SZiFi} master catalogue containing a total of 1499 detections. We cross-match the master catalogue with several external tSZ and X-ray cluster catalogues, setting a lower bound on the purity of our baseline catalogue of 95\,\% and 99\,\% at a minimum signal-to-noise of 5 and 6, respectively. We validate our cluster detection pipeline by applying it to synthetic observations, recovering cluster number counts for which we are able to produce a theoretical prediction that accurately describes them. This validation exercise indicates that our catalogues are well-suited for cosmological inference. The \textit{Planck} \texttt{SZiFi} master catalogue will become publicly available at \href{https://github.com/inigozubeldia/szifi/tree/main/planck_szifi_master_catalogue}{this link}.
\end{abstract}




\section{Introduction}

Galaxy clusters detected through the thermal Sunyaev-Zeldovich (tSZ) effect (\citealt{Sunyaev1972}; see \citealt{Carlstrom2002} and \citealt{Mroczkowski2019} for reviews) are a powerful cosmological probe from which constraints on cosmological parameters such as the matter density parameter $\Omega_{\mathrm{m}}$, the amplitude of matter clustering $\sigma_8$, the equation of state of dark energy, and the sum of the neutrino masses can be obtained (e.g., \citealt{Battye:2003bm,Allen2011,Weinberg13}). Over the past 15 years, a number of galaxy cluster catalogues with $\sim10^2$--$10^3$ objects have been constructed using data from the \textit{Planck} experiment, the Atacama Cosmology Telescope (ACT), and the South Pole Telescope (SPT) (e.g., \citealt{Vanderlinde2010,Hasselfield2013,Planck2014,Bleem2015,Planck2016xxvii,Tarrio2019,Bleem2020,Aghanim2019,Hilton2020, Melin2021,Bleem2022,Bleem2024,Klein2023}). Some of these catalogues have been subsequently used in cosmological number-count analyses (e.g., \citealt{Vanderlinde2010,Hasselfield2013,Planck2014XX,Bleem2015,Ade2016,Bocquet2018,Salvati2018,Zubeldia2019,Salvati2022,Chaubal2022,Bocquet2024,Lee2024}).

With their large sky coverage and broad frequency range, the \textit{Planck} temperature maps remain a competitive dataset for tSZ cluster finding, allowing for the probing of lower cluster masses at low redshifts compared to what is enabled by ACT and SPT (see, e.g., Figure\,9 of \citealt{Bleem2024}). The \textit{Planck} Collaboration released a series of three cluster catalogues: the \textit{Planck} ESZ, PSZ1, and PSZ2 catalogues \citep{Planck2012I,Planck2013XXIX,Planck2016xxvii}, which have been widely used in the literature and employed in several cosmological number-count analyses (e.g., \citealt{Planck2014XX,Ade2016,Salvati2018,Zubeldia2019}). These catalogues, however, were constructed using all the \textit{Planck} High Frequency Instrument (HFI) frequency channels without an assessment of contamination from correlated foregrounds. These foregrounds include the Cosmic Infrared Background (CIB), which has been shown to have the potential to introduce biases in the cluster tSZ observables (see \citealt{Zubeldia2023}). In addition, the approach used to estimate the noise covariance in the cluster finding process has been shown to lead to a bias in the cluster tSZ observables (the `covariance bias') and to a loss of signal-to-noise \citep{Zubeldia2022}. Finally, since \textit{Planck} probes the most massive, nearby clusters, modifications to the tSZ signal from relativistic temperature corrections \citep{Challinor1998, Sazonov1998, Itoh1998, Chluba2012SZpack} may be relevant \citep[e.g.,][]{Remazeilles2019, Rotti2021Giants, Perrott2024}. These relativistic corrections were, however, ignored in the official \textit{Planck} analyses.

In this work, we introduce the \textit{Planck} \texttt{SZiFi} catalogues, a new set of \textit{Planck} cluster catalogues constructed to address these issues and provide a \textit{Planck} galaxy cluster sample produced independently from the official ones. For cluster finding, we use the \textit{Planck} PR3 maps, as opposed to the PR2 ones used for the last official \textit{Planck} catalogue. We search for clusters using \texttt{SZiFi}\footnote{\texttt{\href{https://github.com/inigozubeldia/szifi/}{github.com/inigozubeldia/szifi}}}, the Sunyaev-Zeldovich iterative Finder, a publicly available enhanced implementation of the Multi-frequency Matched Filter (MMF) cluster finding algorithm that has been extensively tested with \textit{Planck}-like synthetic observations \citep{Zubeldia2022,Zubeldia2023}. We detect cluster candidates down to a signal-to-noise threshold of 5. We estimate the MMF noise covariance iteratively, removing the detections from the noise covariance estimate. This procedure eliminates the covariance bias from the cluster tSZ observables and boosts the signal-to-noise to the correct level. In addition, we carefully assess the impact of CIB contamination by (i) obtaining catalogues for several combinations of the \textit{Planck} HFI frequency channels and (ii) employing, for the first time on real data, the spectrally constrained MMFs of \citet{Zubeldia2023}, which we apply in combination with the moment expansion approach of \citet{Chluba2013}. We produce a total of 10 catalogues, each obtained with a different frequency and/or CIB deprojection combination, and compile them into a single \textit{Planck} \texttt{SZiFi} master catalogue. We then cross-match the master catalogue with several existing tSZ and X-ray cluster catalogues containing significantly more clusters than were available when the official \textit{Planck} catalogues were published. Furthermore, for the first time in the context of tSZ cluster detection, we quantify the impact of the relativistic corrections to the tSZ signal on the cluster tSZ observables. Finally, we validate our cluster detection pipeline by applying it to synthetic \textit{Planck}-like observations. 

This paper is structured as follows. First, in Section\,\ref{sec:data} we present the datasets that we use for both cluster detection and the confirmation of cluster candidates. Then, in Sections\,\ref{sec:detection} and\,\ref{sec:confirmation} we describe, respectively, our cluster detection and confirmation pipelines. Next, in Section\,\ref{sec:catalogue} we discuss the properties of our \textit{Planck} \texttt{SZiFi} cluster catalogues, describing the cluster sky position, signal-to-noise and redshift distributions, the purity of the catalogues, and the impact of iterative noise covariance estimation. We then compare one of our catalogues with the official \textit{Planck} MMF3 catalogue in Section\,\ref{sec:comparison} and discuss the impact of the CIB and of the tSZ relativistic corrections in Sections\,\ref{sec:cib} and \ref{sec:rsz}, respectively. Finally, in Section\,\ref{sec:validation} we validate our cluster detection pipeline with synthetic observations and we conclude in Section\,\ref{sec:summary}. In a series of appendices, we offer a brief description of the entries in the \textit{Planck} \texttt{SZiFi} master catalogue (Appendix\,\ref{appendix:catalogue}) and further discussion on the validation of our assessment of CIB contamination (Appendix\,\ref{appendix:mass_selected}) and of our cluster detection pipeline (Appendix\,\ref{appendix:validation}).

\section{Data}\label{sec:data}

\subsection{\textit{Planck} data}

We search for galaxy clusters using the six \textit{Planck} HFI frequency maps, which have bandpasses centred at 100, 143, 217, 353, 545, and 857\,GHz. In particular, we use the 2018 PR3 temperature maps \citep{Planck2020III}, which are publicly available from the Planck Legacy Archive\footnote{\texttt{\href{http://pla.esac.esa.int/pla}{pla.esac.esa.int}}} (PLA).

In our analysis, we apply the \textit{Planck} PR2 Galactic masks leaving 80\% and 70\% of the sky unmasked, the former for cluster detection and the latter imposed in order to obtain the final master catalogue (see Section\,\ref{sec:detection}). We also use the \textit{Planck} PR2 temperature point-source masks for all the six HFI channels, which mask point sources detected at more than 5\,$\sigma$ significance for each channel. All the masks are also available from the PLA.

Finally, in the construction of our matched filters used for cluster detection we use the \textit{Planck} 2018 Reduced Instrument MOdel (RIMO) beams and effective band transmission profiles \citep{Planck2020III}, which are also available from the PLA.

\subsection{Cluster confirmation data}

With the goal of confirming whether our detections correspond to real galaxy clusters and of assigning redshift measurements to them, we cross-match our master catalogue with a number of existing tSZ and X-ray cluster catalogues, which we refer to as our `external catalogues' (see Section\,\ref{sec:confirmation}). We consider the following catalogues and meta-catalogues:

\begin{itemize}

    \item The MCSZ meta-catalogue of tSZ-detected clusters, which is publicly available in the M2C Galaxy Cluster Database\footnote{\texttt{\href{https://www.galaxyclusterdb.eu}{galaxyclusterdb.eu}}}. It contains a total of 2981 objects and was constructed from the \textit{Planck} ESZ, PSZ1, and PSZ2 catalogues \citep{Planck2012I,Planck2013XXIX,Planck2016xxvii}, the South Pole Telescope (SPT) SPT2500 and SPT-ECS catalogues \citep{Bleem2015,Bleem2020}, and the ACT 2008 catalogues of \citet{Marriage2011,Hasselfield2013}. It includes additional validation and redshift information of the \textit{Planck} clusters from a number of post-\textit{Planck} sources \citep{vanderBurg2016,vanderBurg2018,Barrena2018,Streblyanska2018,  Amodeo2018,Burenin2018,Zohren2019, Aguado-Barahona2019,  Streblyanska2019,Zaznobin2019, Boada2019,Barrena2020,Zaznobin2020}.

    \item The MCXC-II meta-catalogue of X-ray-detected clusters (\citealt{Sadibekova2024}, updated from the original MCXC meta-catalogue of \citealt{Piffaretti2011}), also available in the M2C Galaxy Cluster Database. This meta-catalogue contains a total of 2221 objects and is based on the ROSAT All Sky Survey-based catalogues NORAS \citep{Bohringer2000}, REFLEX \citep{Bohringer2000,Chon2012}, REFLEX II \citep{Chon2012}, BCS \citep{Ebeling1998,Ebeling2000}, SGP \citep{Cruddace2002}, NEP \citep{Henry2006}, MACS \citep{Ebeling2007,Mantz2010b,Mann2012,Repp2018}, CIZA \citep{Ebeling2002,Kocevski2007}, and RXGCC \citep{Xu2022}, as well as on the serendipitous catalogues 160SD \citep{Mullis2003}, 400SD \citep{Burenin2007}, SHARC \citep{Burke2003,Romer2000}, WARPS \citep{Perlman2002,Horner2008}, and EMSS \citep{Gioia1994,Henry2004bb}.

    \item The Atacama Cosmology Telescope (ACT) DR5 catalogue of \citet{Hilton2018}\footnote{\texttt{\href{https://lambda.gsfc.nasa.gov/product/act/actpol_dr5_szcluster_catalog_info.html}{lambda.gsfc.nasa.gov/product/act/actpol\_dr5\_szcluster\_\\catalog\_info.html}}}, containing 4195 tSZ detections.
    \item The CODEX catalogue of X-ray detected clusters \citep{Finoguenov2020}. We only consider the subsample with `\texttt{clean=True}', which contains a total of 2815 objects.
    \item The RASS-MCMF catalogue of X-ray detected clusters \citep{Klein2023}. We consider only the 99\,\% purity subsample, which contains a total of 5506 objects and includes revised redshifts for 5 clusters in the PSZ2 catalogue.
    \item The SRG/eROSITA All-Sky Survey (eRASS) catalogue of X-ray detected clusters of \citet{Bulbul2024}. We consider only the cosmology subsample, which has an estimated 95\,\% purity and contains 5259 objects.
    
\end{itemize}

Furthermore, in order to assess the presence of spurious detections in our master catalogue, we cross-match it with the Second \textit{Planck} Catalogue of Compact Sources \citep{Planck2015XXVI} and with the \textit{Planck} Catalogue of Galactic cold clumps \citep{Planck2015XXVIII}, both of which are available from the PLA. 

\section{Cluster detection}\label{sec:detection}

\subsection{Overview}\label{subsec:detection_overview}

We detect galaxy clusters blindly by applying the tSZ cluster finder \texttt{SZiFi} to the \textit{Planck} PR3 HFI frequency maps. \texttt{SZiFi} \citep{Zubeldia2022,Zubeldia2023} is a publicly available enhanced implementation of the multi-frequency matched filter (MMF) cluster finding method \citep{Herranz2002,Melin2006}, a method successfully used for cluster finding by the \textit{Planck}, ACT, and SPT collaborations \citep{Vanderlinde2010,Marriage2011,Hasselfield2013,Planck2012I,Planck2013XXIX,Bleem2015,Planck2016xxvii,Bleem2020,Hilton2020,Bleem2022}. \texttt{SZiFi} incorporates several novel features, most notably iterative noise covariance estimation and the ability to spectrally deproject foregrounds, both of which are used in this work. It is described in detail in \citet{Zubeldia2022,Zubeldia2023}, where its performance was extensively tested with \textit{Planck}-like synthetic observations.

Our cluster detection pipeline can be summarised as follows. First, the \textit{Planck} frequency maps are tessellated into a set of non-overlapping 'selection tiles'. Next, for each selection tile, an MMF is constructed and applied to the corresponding frequency map cut-outs (`frequency fields'), producing a `tile catalogue' by selecting the MMF signal-to-noise peaks greater than a threshold of $q_{\mathrm{th}}=4$. This step requires estimating the noise covariance, which is done iteratively, although we also save the detections obtained in the zero iteration, which form a non-iterative tile catalogue. Each tile catalogue includes, for every detection, its signal-to-noise, which we denote by $q_{\mathrm{obs}}$ following the official \textit{Planck} analysis, which can be used as a mass observable (or mass proxy) in a cosmological analysis (e.g., \citealt{Ade2016}). It also includes estimates of the detection Compton-$y$ parameter value at its centre $\hat{y}_0$, angular size $\hat{\theta}_{500}$, and sky coordinates. In a final step, the (iterative and non-iterative) tile catalogues are merged into a single (iterative and non-iterative) output catalogue. This detection pipeline is described in detail in Section\,\ref{subsec:pipeline}.

In order to assess the impact of contamination due to signals correlated with the cluster tSZ signal (in particular, the CIB), we consider a set of 10 different MMFs formed by various combinations of frequency channels and spectral deprojections. The choices of MMFs and the Spectral Energy Distributions (SEDs) that are deprojected are described in Section\,\ref{subsec:matched_filters_used}. For each MMF, the detection pipeline is applied independently, producing one iterative and one non-iterative catalogue per MMF. These individual catalogues are then mutually cross-matched and compiled into a single `master catalogue' in a final step, as described in Section\,\ref{subsec:master_catalogue}. The master catalogue is then cross-matched against external catalogues, as explained in Section\,\ref{sec:confirmation}. It is important to note, however, that each of the catalogues in the master catalogue should be regarded an independent catalogue in its own right, with a well-defined selection.

\subsection{Matched filters used and CIB deprojection}\label{subsec:matched_filters_used}

Any signal that is spatially correlated with the cluster tSZ signal has the potential to bias the cluster tSZ observables delivered by our cluster detection pipeline (e.g., $q_{\mathrm{obs}}$ and $\hat{y}_0$), and therefore the cluster number counts. This bias can, in turn, potentially propagate to number-count cosmological constraints (see, e.g., \citealt{Zubeldia2023}).

The CIB, formed by the cumulative dust emission from star-forming galaxies (see, e.g., \citealt{Lagache2005}), is a prominent example of an extragalactic signal that is spatially correlated with the tSZ field. This correlation has indeed been measured and has been shown to bias measurements of the tSZ power spectrum and of the cross-correlation of the tSZ field with other tracers of the density field (e.g., \citealt{Planck2013XI,Planck2014XXX,Planck2016XXIII, Hurier2017, Pandey2019, Yan2019, Bleem2022, Sanchez2023, McCarthy2024, McCarthy2024b, Coulton2024}). Dust emission has also been measured at the location of galaxy clusters in stacking analyses using IRAS and \textit{Planck} data \citep{Montier2005, Giard2008, Planck2016XLIII, Planck2016XXIII, Melin2018}. The contribution to the CIB that is spatially correlated with a given cluster sample (the `cluster-correlated CIB') therefore has the potential to bias the cluster tSZ observables. 

This possibility has been discussed in two detailed analyses: \citet{Melin2018} and \citet{Zubeldia2023}. In \citet{Melin2018}, clusters with dust emission from a model calibrated to \textit{Planck} data were injected into \textit{Planck} data and used to assess the impact on the retrieved cluster number counts and cosmological constraints. The impact was found to be small, with about 10\,\% fewer detections in the number counts between $z=0.4$ and $z=0.8$ relative to the case with no dust contamination and negligible changes in the derived cosmological parameters. On the other hand, \citet{Zubeldia2023} performed a simulation-based analysis, using \textit{Planck}-like synthetic maps from the Websky simulation \citep{Stein2019, Stein2020}. They reported a more significant impact on the cluster tSZ observables and, consequently, on the retrieved cluster number counts, suggesting that further analysis of the \textit{Planck} data may be required. 

In order to assess the potential contamination of the cluster observables due to the cluster-correlated CIB, following \citet{Zubeldia2023} we consider a set of different MMFs constructed with various combinations of frequency channels and spectral deprojections, the latter achieved through the use of spectrally constrained MMFs. The choice of channel combinations is informed by the fact that, at the \textit{Planck} frequencies, the CIB signal strongly increases with frequency. Based on this, we consider three frequency combinations: all six HFI channels, the five lowest-frequency channels, and the four lowest-frequency channels. Regarding spectral deprojection, as demonstrated in \citet{Zubeldia2023}, spectrally constrained MMFs, which zero or `deproject' the contribution of one or several foregrounds with given SEDs, can be highly effective at suppressing the CIB-induced bias from the cluster tSZ observables and the cluster number counts. These MMFs are particularly powerful in combination with the moment expansion approach to first order \citep{Chluba2013, Chluba2017, Rotti2021, Vacher2022a}. In this approach, an estimate for the foreground SED is deprojected together with its derivative with respect to one or several parameters parameterising it. Doing this mitigates the impact of assuming a deprojection SED that does not perfectly match the true foreground SED, as shown in \citet{Zubeldia2023} for \textit{Planck}-like synthetic observations (see, in particular, their Figure\,9). This is particularly important for the CIB, as its SED is relatively poorly constrained and varies from cluster to cluster and, on average, as a function of redshift.

\begin{table}
\centering
\begin{tabular}{lll}
\hline

\textbf{MMF} & \textbf{Freq. channels} & \textbf{Spec. deprojection}
\\
\hline

iMMF6 & 100--857\,GHz (6) & None \\
iMMF5 & 100--545\,GHz (5) & None \\
iMMF4 & 100--353\,GHz (4) & None \\

sciMMF6 & 100--857\,GHz (6) & CIB \\
sciMMF5 & 100--545\,GHz (5) & CIB \\
sciMMF4 & 100--353\,GHz (4) & CIB \\

sciMMF6 $\beta$ & 100--857\,GHz (6) & CIB + moment w.r.t. $\beta$ \\
sciMMF6 $\beta_T$ & 100--857\,GHz (6) & CIB + moment w.r.t. $\beta_T$ \\

sciMMF5 $\beta$ & 100--545\,GHz (5) & CIB + moment w.r.t. $\beta$ \\
sciMMF5 $\beta_T$ & 100--545\,GHz (5) & CIB + moment w.r.t. $\beta_T$ \\

\hline
\end{tabular}
\caption{The 10 iterative multifrequency matched filters (MMFs) used in this work with their corresponding frequency channel and spectral deprojection combination. We obtain a cluster catalogue with each of these MMFs.}
\label{table:mmfs}
\end{table}

Taking these considerations into account, we use the following 10 MMFs, all of which are constructed with iterative noise covariance estimation (i.e., they are `iterative', hence the letter `i' in the following acronyms):

\begin{itemize}
    \item \textbf{iMMF6}: Standard iterative matched filter (iMMF) using all six HFI channels.
    \item \textbf{iMMF5}: Standard iterative matched filter (iMMF) using the five lowest-frequency HFI channels.  
    \item \textbf{iMMF4}: Standard iterative matched filter (iMMF) using the four lowest-frequency HFI channels.  
    \item \textbf{sciMMF6}: Spectrally constrained iterative matched filter (sciMMF) using all six HFI channels, deprojecting a model for the cluster-correlated CIB SED (see below).
    \item \textbf{sciMMF5}: Spectrally constrained iterative matched filter (sciMMF) using the five lowest-frequency HFI channels, deprojecting the same model for the cluster-correlated CIB SED.
    \item \textbf{sciMMF4}: Spectrally constrained iterative matched filter (sciMMF) using the four lowest-frequency HFI channels, deprojecting the same model for the cluster-correlated CIB SED.
    \item \textbf{sciMMF6 $\bmath{\beta}$}: Spectrally constrained iterative matched filter (sciMMF) using all HFI channels, deprojecting the same model for the cluster-correlated CIB SED and its first-order moment with respect to the SED parameter $\beta$ (see below). 
    \item \textbf{sciMMF6 $\bmath{\beta}_T$}: Spectrally constrained iterative matched filter (sciMMF) using all HFI channels, deprojecting the same model for the cluster-correlated CIB SED and its first-order moment with respect to the SED parameter $\beta_T$ (see below). 
    \item \textbf{sciMMF5 $\bmath{\beta}$}:  As sciMMF6 $\beta$, but using only the five lowest-frequency HFI channels.
    \item \textbf{sciMMF5 $\bmath{\beta}_T$}:  As sciMMF6 $\beta_T$, but using only the five lowest-frequency HFI channels.  
\end{itemize}
A summary of the frequency channel and spectral deprojection combinations of our 10 iterative MMFs can be found in Table\,\ref{table:mmfs}. We stress that each combination leads to its own catalogue of detections. In addition to these iterative MMFs, we also consider their non-iterative counterparts in order to assess the impact of iterative noise covariance estimation and for better comparison with the official \textit{Planck} catalogues.

Following \citet{Planck2014XXX}, we assume that, at the \textit{Planck} frequencies, the CIB SED at a given redshift can be described with a modified blackbody,
\begin{equation}\label{eq:sed}
    I_{\nu} (z) = [\nu (1+z)]^{\beta} B_{\nu (1+z)} [ T_0 (1+z)^{\alpha}],
\end{equation}
where $\nu$ is the observation frequency, $\beta$ is the dust emissivity index, $T_0$ is the dust temperature at $z=0$, $\alpha$ is the dust evolution slope, and $B_\nu (T)$ is Planck's function. We further define the dust temperature parameter as $\beta_T \equiv T_0^{-1}$. 

For each of our sciMMFs, we evaluate the CIB SED and, if relevant, its derivatives with respect to $\beta$ or $\beta_T$, at $\beta=1.75$, $T_0=24.4$\,K, and $\alpha=0.36$. These parameter values are taken to be consistent with their \textit{Planck} constraints in \citet{Planck2014XXX}, which were obtained assuming the same SED model as adopted here. In addition, we evaluate the CIB SED at $z=0.2$, a redshift that is representative of the \textit{Planck} cluster sample (see Section\,\ref{sec:catalogue}). We note, in particular, that even if the CIB peaks at higher redshifts, it is the part that is spatially correlated with our cluster sample that must be removed, as any uncorrelated contribution only adds variance to the cluster observables (see \citealt{Zubeldia2023}). We also note that we deproject the moment with respect to $\beta_T$, rather than $T_0$, as $\beta_T$ has been shown to be a better parametrisation than $T_0$ regarding moment deprojection \citep{Chluba2017}. Finally, in combination with the moment expansion approach, our sciMMFs have been shown to be effectively immune to the choice of the parameter values in the deprojection SED, provided that they are within a reasonable range. In particular, in \citet{Zubeldia2023} it was shown for \textit{Planck}-like synthetic data based on the Websky simulation, in which the CIB has $\beta=1.6$ and $T_0=20.7$\,K, that evaluating $\beta$ across a range from 1.4 to 1.8 resulted in negligible changes in the recovered signal-to-noise values for sciMMF6 $\beta$, remaining unbiased, and in very small changes for sciMMF6 $\beta_T$ (see their Figure\,9). Similarly, evaluating $T_0$ across a range from 16.8\,K to 24.8\,K resulted in negligible changes in the signal-to-noise for both sciMMF6 $\beta_T$ and sciMMF6 $\beta$, both remaining unbiased.

Spectral deprojection and/or the use of fewer frequency channels leads to a signal-to-noise penalty relative to the optimal case, which is given by iMMF6, with fewer and less significant detections: foreground mitigation is, therefore, potentially achieved at a cost. This signal-to-noise penalty is quantified in Section\,\ref{sec:catalogue}.

\subsection{Cluster detection pipeline}\label{subsec:pipeline}

In this section we describe how we use \texttt{SZiFi} to produce each of our catalogues using the \textit{Planck} PR3 HFI maps. We refer the reader to \citet{Zubeldia2022,Zubeldia2023} for further details about the code and, especially, for detailed discussions of its novel elements (iterative noise covariance estimation, foreground spectral deprojection), as well as for tests of its performance with synthetic data.

\subsubsection{Sky tessellation}\label{subsec:tessellation}

\texttt{SZiFi} starts by tessellating the sky into a set of non-overlapping selection tiles, each of which is processed independently. As well as allowing for local noise estimation, with the noise and foregrounds in the \textit{Planck} maps varying significantly across the sky, this procedure also enables for the use of the flat-sky approximation when analysing the maps in harmonic space. We choose the tessellation scheme provided by a HEALPix pixelation \citep{Gorski2005} with $N_{\mathrm{side}}=8$, which leads to a total of 768 tiles, each with an area of 53.7\,deg$^2$ and a typical extent of $53.7^{1/2}$\,deg= 7.3\,deg. This scheme guarantees that the entire sky is covered and that each point in the sky belongs to only one tile, avoiding the possibility of double-counting of pixels.


Taking this tessellation scheme into account, \texttt{SZiFi} extracts square cut-outs (hereafter, `frequency fields') of 14.8\,deg $\times$ 14.8\,deg from the \textit{Planck} HFI frequency maps centred at the centre of each selection tile. These fields are obtained as equirectangular (or \textit{plate carr\'{e}e}) projections using the \texttt{healpy}\footnote{\texttt{\href{healpy.readthedocs.io}{healpy.readthedocs.io}}} library. Each frequency field consists of 1024 $\times$ 1024 pixels with a pixel size of 0.867\,arcmin. The chosen cut-out size ensures that every tile is fully contained within its corresponding field, regardless of its shape. We also obtain cut-outs from two masks (`mask fields'): (i) the \textit{Planck} PR2 80\,\% Galactic mask and (ii) a point-source mask constructed as the union of the \textit{Planck} PR2 point-source masks for all HFI channels. The mask fields have the same size and pixelation as the frequency fields. Taking the Galactic mask into account, there is a total of 627 tiles that remain not fully masked.


In order to suppress spectral leakage when taking Fourier transforms, each Galactic mask field is apodised with the `smooth' scheme of \texttt{pymaster}\footnote{\texttt{\href{namaster.readthedocs.io}{namaster.readthedocs.io}}}, which zeroes all pixels closer to 2.5 times the apodisation scale to a masked pixel, then convolves the resulting mask with a Gaussian kernel with its standard deviation equal to the apodisation scale, and finally zeroes all the originally-masked pixels. We choose an apodisation scale of 12\,arcmin, noting that the edges of the mask fields are also apodised by this procedure.


\subsubsection{Field preprocessing and noise covariance estimation}\label{subsec:preprocess}

Consider a given selection tile. \texttt{SZiFi} then proceeds by applying the corresponding point-source mask to each frequency field, masking bright point sources that may lead to spurious detections. The masked regions are then inpainted with the real-space diffusive inpainting algorithm of \citet{Gruetjen2015}, using a total 100 iterations, after which the apodised Galactic mask is applied. Next, a Fourier-space top-hat filter is imposed to the frequency fields, with a minimum multipole $l_{\mathrm{min}} = 100$ and a maximum multipole $l_{\mathrm{max}} = 2500$, selecting the scales relevant for cluster finding that are not noise-dominated.

Next, \texttt{SZiFi} estimates the noise cross-channel power spectrum matrix (`noise covariance'), an essential element in an MMF. Here, by noise we mean instrumental noise as well as foreground signals (the CMB, the CIB, the kSZ signal, Galactic foregrounds, etc.). For each tile, this is done with the inpainted, apodised frequency fields using the MASTER algorithm \citep{Hivon2002} as implemented in \texttt{pymaster} \citep{Alonso2019}. The cross power spectra are estimated in a set of 51 equally-spaced bands centred between $l = 48.6$ and $l = 2519.9$. We properly account for the multipole coupling due to the Galactic mask by multiplying each pseudo power spectrum by the inverse of the mask coupling matrix, which is computed with \texttt{pymaster} for each tile.

In this work we use the iterative noise covariance estimation approach introduced in \citet{Zubeldia2022}, with one iteration. In this approach, in the initial iteration, the noise covariance is computed using the input frequency maps, as we just described. Once the first, non-iterative tile catalogue is obtained (as detailed below), the same noise covariance estimation pipeline is applied to frequency maps in which the detections above a signal-to-noise threshold $q_{\mathrm{mask}} = 4$ have been masked out. This procedure leads to an updated, iterative tile catalogue. In Section\,\ref{subsec:iterative} we offer a brief review of the benefits of iterative noise estimation.

\subsubsection{Matched filtering}\label{subsec:mmf}

\begin{figure*}
\centering
\includegraphics[width=0.8\textwidth,trim={0mm 0mm 0mm 0mm},clip]{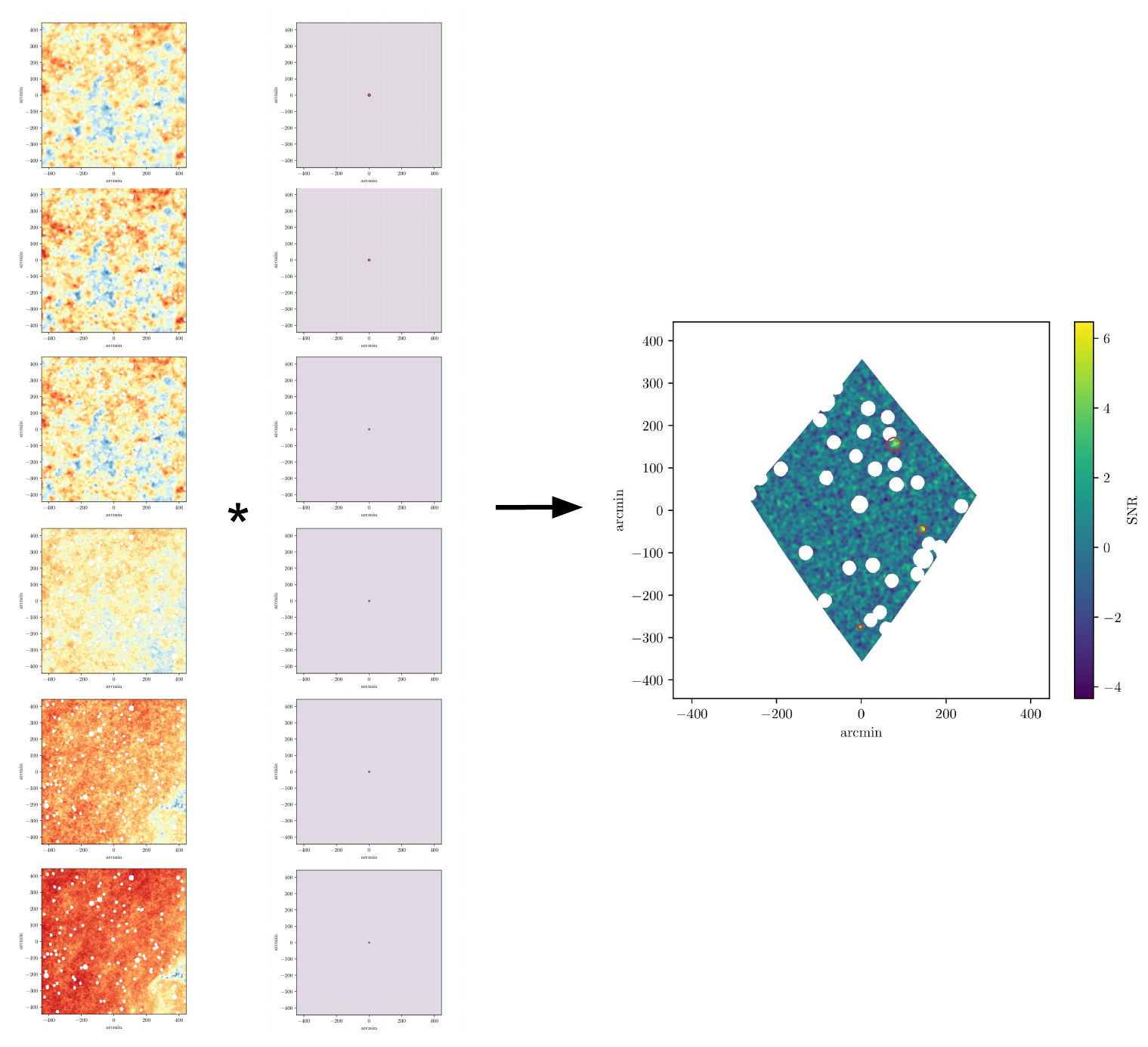}
\caption{Illustration of the MMF cluster detection pipeline as implemented in \texttt{SZiFi} for one of our selection tiles and our iMMF6 filter. The left column shows the six \textit{Planck} HFI frequency fields, from 100\,GHz (top), where the CMB dominates, to 857\,GHz (bottom), where dust becomes dominant. The fields are $14.8 \times 14.8$\,deg across; note the point-source mask. In the MMF, these fields are convolved, in an optimal way, with the cluster template for each channel for a set of search angular scales $\theta_{500}$. Here, the middle column shows the template for $\theta_{500} = 5$\,arcmin. The output of this matched filtering operation is a signal-to-noise map for each angular scale, shown on the right for $\theta_{500} = 5$\,arcmin and after the selection mask has been applied. Note that this mask includes the tile mask that corresponds to a HEALPix tessellation with $N_{\mathrm{side}} = 8$ and prevents double-counting of pixels. The detections for this tile (three of them) are shown as the red circles, with their radius corresponding to the angular scale at which they are detected.}
\label{fig:pipeline}
\end{figure*}

For each tile, once the noise covariance is estimated \texttt{SZiFi} constructs an MMF for each of a set of `search' angular scales, assuming (i) a spatial and a spectral template for the expected cluster tSZ signal and (ii) a spectral template (SED) to be deprojected, if relevant.

More precisely, an MMF assumes that the frequency fields at pixel $\mathbfit{x}$, $ \mathbfit{d} (\mathbfit{x})$, can be written as
\begin{equation}
    \mathbfit{d} (\mathbfit{x}) = \mathbfit{y} (\mathbfit{x}) + \mathbfit{n} (\mathbfit{x}).
\end{equation}
Here, $\mathbfit{y} (\mathbfit{x})$ is the signal that is being targeted, i.e., the galaxy cluster tSZ signal. It is assumed that its $i$-th component can be written as $y_i (\mathbfit{x}) = a_{i} (y \ast b_i )(\mathbfit{x})$, where $a_{i}$ is the tSZ SED at frequency channel $i$, $y(\mathbfit{x})$ is the Compton-$y$ map describing the spatial variation of the tSZ signal, $b_i$ is the instrument beam at frequency channel $i$ and $\ast$ denotes convolution. In addition, $\mathbfit{n} (\mathbfit{x})$ is some additive `noise' that also includes all the other components in the maps. These components (the CMB, the kinetic Sunyaev-Zeldovich signal, the CIB, Galactic foregrounds, etc.) may or may not be correlated with the tSZ signal. It is further assumed that, for a single galaxy cluster, its Compton-$y$ map $y (\mathbfit{x})$ can be written as $y (\mathbfit{x}) = y_0 y_{\mathrm{t}} (\mathbfit{x} ;  \theta_{500}, \bmath{\theta}_\mathrm{c} )$, where $y_0$ is an amplitude parameter, which we take to be the value of the cluster's Compton-$y$ parameter at the cluster centre, and $y_{\mathrm{t}} (\mathbfit{x} ; \theta_{500}, \bmath{\theta}_\mathrm{c})$ is a spatial template. This template depends on the cluster angular size $\theta_{500}$, defined, as is customary, as the angle subtended by $R_{500}$, the radius within which the mean density is equal to 500 times the critical density at the cluster's redshift. The template also depends on the sky coordinates of the cluster centre, $\bmath{\theta}_\mathrm{c}$. 

At given input values for $\theta_{500}$ and $\bmath{\theta}_\mathrm{c}$, a general MMF estimator for $y_0$, spectrally deprojecting $N_{\mathrm{dep}}$ components, can then be written, assuming the flat-sky approximation, as \citep{Zubeldia2023} 
\begin{multline}
\hat{y}_{0} (\theta_{500},\bmath{\theta}_\mathrm{c}) =  \\ N^{-1} \int  \frac{ d^2 \mathbfit{l}}{2 \pi}  \, \frac{y_{\mathrm{t}}^{\ast} (\theta_{500},\bmath{\theta}_\mathrm{c})   \mathbfit{c}^T (\mathbfss{A}^T \mathbfss{C}^{-1} \mathbfss{A})^{-1} \mathbfss{A}^T  \mathbfss{C}^{-1}  \mathbfit{d} }{\mathbfit{c}^T (\mathbfss{A}^T \mathbfss{C}^{-1} \mathbfss{A})^{-1} \mathbfit{c}},
\end{multline}
where
\begin{equation}
   N =  \int \frac{ d^2 \mathbfit{l}}{2 \pi} \frac{ y_{\mathrm{t}}^{\ast}  y_{\mathrm{t}}  }{\mathbfit{c}^T (\mathbfss{A}^T \mathbfss{C}^{-1} \mathbfss{A})^{-1} \mathbfit{c}},
\end{equation}
and where we have left the dependencies of $\mathbfit{d}$, $y_{\mathrm{t}}$, and $\mathbfss{C}$ on multipole $\mathbfit{l}$ implicit to avoid clutter in the notation. Here, $\mathbfss{C}$ is the noise covariance matrix; $\mathbfss{A} = [\mathbfit{a} \,\, \mathbfit{b}_1 \dots \mathbfit{b}_{N_{\mathrm{dep}}} ] $ is a $N_{\mathrm{f}} \times N_{\mathrm{dep}} + 1$ mixing matrix where $N_{\mathrm{f}}$ is the number of frequency channels, $\mathbfit{a}$ is the tSZ SED, and $\mathbfit{b}_i$ is the SED of the i-th component to be deprojected, noting that each row in the matrix is rescaled for each multipole $\mathbfit{l}$ by the Fourier transform of the beam; $\mathbfit{c}^T = [1 \,\, 0 \dots 0]$ is a vector of dimension $N_{\mathrm{dep}} + 1$ that ensures that the correct component (the tSZ signal) is retrieved; and $y_{\mathrm{t}} (\mathbfit{l} ; \theta_{500}, \bmath{\theta}_\mathrm{c})$ is the Fourier transform of the cluster spatial template. Note that in the absence of any spectral deprojection, $\mathbfss{A} = [\mathbfit{a}]$ and this general MMF reduces to the standard MMF \citep{Herranz2002,Melin2006} used widely for cluster detection.

The variance of this general MMF estimator is given by $\sigma_{\mathrm{f}}(\theta_{500}) = N^{-1}$ \citep{Zubeldia2023}, which allows to define the signal-to-noise as
\begin{equation}\label{eq:snr_def}
q (\theta_{500}, \bmath{\theta}_{\mathrm{c}}) \equiv \frac{\hat{y}_{0} (\theta_{500},\bmath{\theta}_\mathrm{c})}{\sigma_{\mathrm{f}}(\theta_{500})}.
\end{equation}
At a given angular scale $\theta_{500}$, a map of $\hat{y}_0$ and $q$ estimates can be produced by placing the template at the centre of every pixel. This operation can be thought of as a convolution, and it is implemented as such in \texttt{SZiFi}, boosting significantly the  computational performance (see \citealt{Zubeldia2022} for more details).

As in \citet{Planck2015XXVIII},  we take the cluster spatial template to be the Compton-$y$ signal due to the \citet{Arnaud2010} pressure profile, setting the concentration to $c_{500} = c = 1.177$ also following \citet{Planck2015XXVIII}. \texttt{SZiFi} integrates the pressure profile numerically up to a radius of $5 R_{500}$, at which it is truncated (as in \citealt{Planck2015XXVIII}). For the spectral template $\mathbfit{a}$, we assume the non-relativistic tSZ SED, integrated within each frequency channel weighted by the channel's effective band transmission profile. When deprojecting the CIB and its moments, we assume the modified blackbody SED described in Section\,\ref{subsec:matched_filters_used}, also integrated for each channel weighted by the effective band transmission profile.

We produce $\hat{y}_0$ and $q$ maps at a set of 25 search angular scales logarithmically spaced between $\theta_{500} = 0.5$\,arcmin and $\theta_{500}=32$\,arcmin. The use of several angular scales allows to optimise the detection of clusters with different angular sizes. The chosen range of angular scales is the same as that used in \citet{Planck2015XXVIII} and is designed to span the angular sizes of most clusters that are expected to be detected by \textit{Planck}.

In summary, the matched filtering step produces, for each selection tile, a three-dimensional distribution for $\hat{y}_0$ and $q$,  $\hat{y}_0 (\mathbfit{x} ;  \theta_{500})$ and $q (\mathbfit{x} ;  \theta_{500})$, spanning the two sky coordinates $\mathbfit{x}$ and the angular scale $\theta_{500}$. We note that \texttt{SZiFi} also supports extracting $\hat{y}_0$ and $q$ at a set of input angular scales and sky positions via what is known as its fixed mode. We make use of this functionality in Sections\,\ref{sec:cib} and\,\ref{sec:rsz}.


\subsubsection{Peak finding}\label{subsec:peak}

Next, \texttt{SZiFi} turns the three-dimensional signal-to-noise distribution into a catalogue of cluster candidates for each tile. In order to achieve this, first, for each angular scale, the signal-to-noise map is masked with a `peak-finding mask' that is defined as the union of the unapodised Galactic mask, the point-source mask, and the tessellation mask, the latter being a binary mask zeroing all the pixels falling outside the corresponding HEALPix tile. The tessellation mask ensures that each point in the sky is only covered once for the purposes of cluster detection, avoiding double-counting of pixels. We note that, thanks to the nature of our tessellation scheme, the tessellation mask ensures that all the pixels falling within it are far away from the edges of the square field used for the construction of the MMF (see Figure\,\ref{fig:pipeline}). This suppresses the bias in the MMF observables that can arise in detections near the edges. An alternative solution to this problem was followed in \citet{Planck2016xxvii}, in which the original tessellation of the sky was updated by recentering the tiles at the locations of the detected cluster candidates.

A first tile catalogue of cluster candidates is then formed by the local peaks of the three-dimensional masked signal-to-noise distribution $q (\mathbfit{x} ;  \theta_{500})$ that are above a selection threshold $q_{\mathrm{th}}$, which we choose to be $q_{\mathrm{th}} = 4$. These local peaks are found with a real-space peak-finding filter that identifies a peak as a (three-dimensional) pixel with a signal-to-noise value larger than all its neighbouring pixels. We denote the signal-to-noise of each detection with $q_{\mathrm{obs}}$. The corresponding values of $\hat{y}_0$ and of the filter angular scale $\theta_{500}$ are assigned as the detection estimates for $y_0$ and $\theta_{500}$, and the pixel coordinates of the peak as the sky coordinates of the detection. A final `selection mask' is then imposed on the detected cluster candidates. This selection mask is a union of two masks. The first one is the union of the unapodised Galactic mask with the point-source mask increased by zeroing all the pixels closer to masked pixels by a buffer distance of 10\,arcmin. This buffering procedure minimises the chances of spurious detections close to the mask edges due to ringing. The second mask is the tessellation mask.

As discussed in \citet{Zubeldia2021}, these `blind' measurements of $y_0$, $\theta_{500}$ and the sky position can be seen as maximum-likelihood estimates and are therefore, in general, biased (except the two sky coordinates, due to symmetry). In addition, for each cluster, the blind or `optimal' signal-to-noise $q_{\mathrm{obs}}$ is biased high relative to the signal-to-noise that would be extracted at the cluster's true sky position and angular size, as it is obtained via a maximisation procedure over noisy data. This `optimisation bias' was first discussed in \citet{Vanderlinde2010} and subsequently studied in detail in \citet{Zubeldia2021}. As shown, e.g., in \citet{Zubeldia2021}, it can be modelled with a simple prescription, which we follow in the theoretical description of our validation catalogues in Section\,\ref{sec:validation}.

Figure\,\ref{fig:pipeline} illustrates our matched filtering and peak finding process for a reference selection tile (with a HEALPix index of 13) for our iMMF6 filter. The maps on the left correspond to the point-source-masked frequency fields shown, from top to bottom, in increasing frequency. They are convolved by the MMF template at $\theta_{500} = 5$\,arcmin (maps in the middle of the figure), producing a signal-to-noise map (rightmost map) that we show masked by the corresponding selection mask. Note, in particular, the shape of the HEALPix selection tile and how the point-source-masked region has increased in size. In this map detections above a signal-to-noise threshold of 5 (three of them) are shown as the red circles, with the radius of each circle corresponding to the filter angular scale at which the detection was found. We note that this signal-to-noise map corresponds to the second (and final) iteration of the noise covariance estimation.
 
\subsubsection{Iterative noise covariance estimation}\label{subsec:iterative}

Once a first, non-iterative tile catalogue is constructed, \texttt{SZiFi} proceeds by producing a `cluster mask' in which all the pixels falling within $3 \hat{\theta}_{500}$\,arcmin of the estimated centre of each cluster candidate with $q_{\mathrm{opt}}$ above a masking threshold are zeroed. We choose this threshold to be $q_{\mathrm{mask}} = 4$, i.e., equal to the detection threshold. We note that in the construction of the cluster mask, \texttt{SZiFi} also includes cluster candidates which fall within the buffered Galactic + point-source mask, but outside the tessellation mask. This is done because these detections also contribute to the noise covariance estimate, as the tessellation mask is only imposed in the peak-finding step.

After computing the cluster mask, \texttt{SZiFi} multiplies it by the point-source mask, leading to a new `point source' mask. The pipeline is then re-run from the preprocessing step of Section \ref{subsec:preprocess} using the new cluster + point-source mask instead of the original point-source mask in the covariance estimation step. This procedure generates a new noise covariance estimate, which is then used to construct a new set of MMFs that in turn lead to an updated, iterative tile catalogue. We note that the original peak-finding and selection masks are used: the cluster candidates are only masked out for noise estimation purposes. In order to allow for better comparison with the official \textit{Planck} catalogues we also save each non-iterative tile catalogue (see Section\,\ref{sec:comparison}).

As shown in \citet{Zubeldia2022}, iterative noise covariance estimation solves two different issues that arise if the noise covariance estimate is simply taken to be the covariance of the frequency maps (our first step in the iterative approach), both of which arise due to the presence of the tSZ signal in what is regarded as the noise in the data. First, if the tSZ signal of the detections is present in the maps used for covariance estimation, the noise covariance is overestimated. This translates into a loss of signal-to-noise, with fewer and less significant detections. Second, a negative bias in the tSZ observables ($q_{\mathrm{obs}}$ and $\hat{y}_0$) arises due to the presence of the signal in the (noisy) covariance estimate that is obtained from the same data on which the MMF is subsequently applied. This `covariance bias' is analogous to the so-called ILC bias that is encountered in CMB component separation with Internal Linear Combinations (ILC) methods (e.g., \citealt{Delabrouille2009}; see \citealt{Coulton2024} for a bias mitigation approach in the context of power spectrum estimation with parallels to our masking algorithm). As discussed in \citet{Zubeldia2022}, iterative noise covariance estimation with just one iteration effectively removes both issues, boosting the detection signal-to-noise to the correct level and eliminating the covariance bias. We note that the official \textit{Planck} catalogues were obtained without iterative noise covariance estimation, simply taking the noise covariance to be equal to the covariance of the data (see, e.g., \citealt{Planck2016xxvii}). 

\subsubsection{Tile catalogue merging}\label{subsec:merging}

Once the iterative tile catalogues are produced, \texttt{SZiFi} merges them into a single survey-wide catalogue for each MMF, assigning them global sky coordinates (Galactic longitude and latitude, as well as equatorial coordinates).

As a final step, \texttt{SZiFi} merges detections lying within a given angular distance of each other into a single one. This merging step solves two problems. First, our peak-finding algorithm, described in Section\,\ref{subsec:peak}, sometimes leads to several detections located close to each other that are associated to a single true object, as it finds local peaks in the signal-to-noise three-dimensional distribution, which is noisy. In addition, clusters lying very close to the border between two selection tiles may be detected in both tiles (or in more than two, if the cluster is near the intersection between four tiles). 

The following merging algorithm is applied. First, the detection with the highest signal-to-noise in the full catalogue is identified. All the detections falling within an angular distance of its estimated angular scale $\hat{\theta}_{500}$ are considered as belonging to the same object and dropped from the catalogue. This process is repeated until the lowest signal-to-noise detection in the catalogue is reached. Apart from a few cases (see Section\,\ref{subsec:inspection} and Section\,\ref{subec:purity_sim}), this algorithm successfully merges all detections belonging to the same object. In addition, as we show in Section\,\ref{sec:validation} by applying \texttt{SZiFi} to our synthetic \textit{Planck}-like maps, this algorithm leads to a cluster catalogue that we can successfully model theoretically. This would not be the case if, e.g., it led to many instances of merging of detections corresponding to different clusters.

This merging algorithm is also applied to the non-iterative tile catalogues, producing survey-wide non-iterative catalogues. Hereafter, we will refer to these catalogues using the same naming convention used for their iterative counterparts but dropping the initial `i' (e.g., MMF6 instead of iMMF6).

Our merging algorithm is similar to the one used for the construction of the official \textit{Planck} catalogues, although a fixed merging angular separation of 10\,arcmin was employed then (e.g., \citealt{Planck2015XXVIII}). It differs significantly, however, from the friends-of-friends merging algorithm used in \citet{Zubeldia2022,Zubeldia2023}. We have verified that applying this friends-of-friends merging algorithm with a linking length of $\theta_{\mathrm{id}} =10$\,arcmin to our catalogues leads to a very small ($\mathcal{O}(1)$) change in the number of clusters in the output catalogues.

After this merging step, we impose a new selection threshold of $q_{\mathrm{th}} = 5$. 

\subsection{Construction of the master catalogue}\label{subsec:master_catalogue}

After obtaining our 10 iterative catalogues of cluster candidates and their 10 non-iterative counterparts, we mutually cross-match them against each other and compile them into a single `master catalogue', assigning a unique name to each detection. This master catalogue is designed to allow for the retrieval of each individual catalogue while making it easy to access the cluster observable values for each detection for all the filters with which it was detected.

\subsubsection{Cross-matching algorithm}\label{subsec:crossmatch}

The construction of the master catalogue is based on the following cross-matching algorithm. Consider two catalogues, catalogue $A$ and catalogue $B$. The algorithm starts by finding the object in catalogue $A$ with the largest value of the signal-to-noise, which we denote with $a$. It then finds all the objects in catalogue $B$ lying within a given cross-matching angular distance $\theta_{\mathrm{match}}$ of $a$. If there is only one such object, that object is identified as the same object as $a$, and is removed from catalogue $B$. If there are multiple such objects, the object with the largest value of the signal-to-noise is identified as the same object as $a$, and similarly removed from catalogue $B$. If there are none, $a$ is labelled as having no counterpart in catalogue $B$. In any of the three cases, $a$ is then removed from catalogue $A$. The process is then repeated until the last object in catalogue $A$ has been reached. In the last step, any remaining objects in catalogue $B$ are labelled as having no counterpart in catalogue $A$. 

This cross-matching algorithm returns two cross-matched catalogues with the same number of objects in them, $A_{\mathrm{C}}$ and $B_{\mathrm{C}}$. In them, each entry contains the original entries in $A$ and $B$, respectively, for each cross-matched object, and a null entry in either $A_{\mathrm{C}}$ or $B_{\mathrm{C}}$ if the given object has not been found a counterpart in $A$ or $B$, respectively.

We adopt a cross-matching distance of $\theta_{\mathrm{match}} = 10$\,arcmin, which is comparable with the \textit{Planck} HFI beams. Given the low number density of our detections in the sky, we argue that this simple cross-matching algorithm is well-suited for our catalogues. Indeed, our iMMF6 catalogue within the cosmology mask (see below for how the mask is defined), which contains the largest number of detections, features 833 detections over 58.65\,\% of the sky (see below), with 29.04\,deg$^2$ per cluster on average. We also argue that our algorithm is similarly well-suited for cross-matching with the external catalogues (see Section\,\ref{sec:confirmation}), for which the number density is, on average, at most $\sim 10$ times larger than in our iMMF6 catalogue.

\subsubsection{Master catalogue}

In order to construct the master catalogue, we start by cross-matching the iMMF6 catalogue, which features the largest number of detections, with its non-iterative counterpart, the MMF6 catalogue, creating a first master catalogue containing both the successful cross-matches and the remaining detections. We then sequentially cross-match the resulting master catalogue with the remaining catalogues, in the following order, with the non-iterative catalogues cross-matched immediately after their iterative counterparts: sciMMF6, iMMF5, sciMMF5, sciMMF6 $\beta_T$, iMMF4, sciMMF4, sciMMF6 $\beta$, sciMMF5 $\beta_T$, sciMMF5 $\beta$. 

The master catalogue is formed by a total of 1499 detections. Each detection in it contains entries for the cluster observables ($q_{\mathrm{obs}}$, $\hat{y}_{0}$, $\hat{\theta}_{500}$, and sky coordinates) for \emph{all} 10 iterative and 10 non-iterative catalogues. Each entry contains the corresponding value for all the catalogues in which the detection is found, and a null value otherwise. Each detection also contains an entry for its `master sky coordinates', which are the detection sky coordinates for the first catalogue in our cross-matching procedure in which the detection appears.

After constructing the master catalogue, we apply a more restrictive Galactic mask to it, the \textit{Planck} PR2 70\,\% Galactic mask, flagging all the detections falling outside of it. We refer to the union of this mask with the buffered Galactic+point-source mask that was employed for cluster finding as the `cosmology mask'. It leaves 58.65\,\% of the sky unmasked and can be seen as the grey area in Figure\,\ref{fig:map_detections}. Within it, the master catalogue contains a total of 1152 detections. In the following (Sections\,\ref{sec:catalogue}--\ref{sec:cib}), we will always consider our catalogues restricted to the area left unmasked by the cosmology mask.

As already noted, we construct this master catalogue for convenience, as it allows for easy access to the cluster observables obtained with different filters for objects detected with more than one filter, as well as for easy cross-matching with the external catalogues. However, each of our 10 iterative catalogues, as well as each of their non-iterative counterparts, must be thought of as an independent catalogue in its own right, with a well-defined tSZ selection. Each catalogue remains unaffected by the cross-matching algorithm and can be fully retrieved from the master catalogue.

\section{Cluster confirmation}\label{sec:confirmation}

\subsection{Cross-matching with the external catalogues and redshift assignment}\label{subsec:cross_match}

We cross-match the master catalogue with the external catalogues described in Section\,\ref{sec:data} with the goal of determining whether each detection in our master catalogue corresponds to a real galaxy cluster and, if so, of assigning a redshift measurement to it. We recall that these are the MCSZ meta-catalogue, the MCXC-II meta-catalogue, the ACT DR5 catalogue, the CODEX catalogue, the RASS-MCMF catalogue, and the eRASS catalogue. We perform the cross-matching using the master sky coordinates of each detection and applying a modified version of the algorithm described in Section\,\ref{subsec:crossmatch}. In particular, in the instances for which there are several objects in a given external catalogue within $10$\,arcmin of one of our detections, we assign the closest object, as opposed to the one with the highest signal-to-noise, as the signal-to-noise is not defined, in general, for the objects in the external catalogues. We assign a cross-matching flag to each successful cross-match in the master catalogue, as well as the name(s) of its counterpart(s) in the external catalogues. Of the 1152 detections in the master catalogue (within the cosmology mask), we find 869 successful cross-matches with the MCSZ meta-catalogue, 486 with the MCXC-II meta-catalogue, 294 with the ACT DR5 catalogue, 286 with the CODEX catalogue, 699 with the RASS-MCMF catalogue, and 309 with the eRASS catalogue. All together, we are able to find positive cross-matches for 957 of the 1152 objects in the master catalogue within the cosmology mask.

We consider all these cross-matched objects in our master catalogue to be real galaxy clusters, except for the objects that have only been cross-matched with an unconfirmed detection in the \textit{Planck} PSZ1 or PSZ2 catalogues, which are contained in the MCSZ meta-catalogue. In particular, in the PSZ1 catalogue, we consider an unconfirmed detection to be an object of class 3, and in the PSZ2 catalogue, an object of class \mbox{-1}. We find 1 and 60 such cross-matches with the PSZ1 and PSZ2 catalogues, respectively. We introduce a confirmation flag that is set to 1 for all objects with a positive cross-match except for these cross-matches with unconfirmed \textit{Planck} detections, and to 0 for the remaining objects. We note that all the cross-matches in the master catalogue with only one object in either any of the SPT catalogues hosted in the MCSZ meta-catalogue or in the ACT DR5 catalogue are cross-matches with a confirmed cluster in those catalogues.

In addition, we assign redshift measurements as follows. For every successful cross-match, if there is only a redshift measurement in one of the external catalogues with which the cross-match has been successful, we assign it as the cluster's redshift. For detections with cross-matches with available redshifts in more than one external catalogue, we adopt the following assignment hierarchy, giving preference to spectroscopic measurements over photometric ones: spec-$z$ in MCSZ, spec-$z$ in MCXC-II, spec-$z$ in RASS-MCMF, spec-$z$ in ACT DR5, spec-$z$ in CODEX, photo-$z$ in MCSZ, photo-$z$ in MCXC-II, photo-$z$ in ACT DR5, photo-$z$ in RASS-MCMF, photo-$z$ in eRASS. We break this hierarchy for four clusters, PSZ2 G109.86+27.94, PSZ2 G181.71-68.65, PSZ2 G281.09-42.51, and PSZ2 G287.00-35.24, in their PSZ2 names, for which updated, significantly different redshift values have been published (see Table\,A1 of \citealt{Klein2023}). We assign these updated redshift values for these four clusters, instead of those in the MCSZ meta-catalogue. Within the cosmology mask, we are able to assign redshift values to 882 out of a total of 1152 detections.

To each cluster with an available redshift measurement we also assign a redshift type flag, identifying the type of redshift measurement (spectroscopic, photometric, or unspecified). Within the cosmology mask, 742 of the redshift measurements in our master catalogue are spectroscopic, 132 photometric, and the remaining 8 of an unspecified type.

\subsection{Cross-matching with \textit{Planck} GCCs and point sources}

Galactic cold clumps (GCCs) can be detected by tSZ cluster detection methods, appearing as spurious detections (see, e.g., \citealt{Planck2016xxvii}). In order to assess the contamination of the master catalogue by GCCs, we cross-match it with the \textit{Planck} Catalogue of cold clumps \citep{Planck2015XXVIII}, which contains 13242 objects. We do this using the same cross-matching algorithm employed to cross-match the master catalogue with the external catalogues (Section\,\ref{subsec:cross_match}). We find 17 cross-matches within the cosmology mask, and flag them in the master catalogue, additionally setting their confirmation flag to -1, indicating that they are likely false detections.

Moreover, bright millimetre point sources can also be spuriously detected as tSZ detections. We further cross-match the master catalogue with the Second \textit{Planck} catalogue of Compact Sources \citep{Planck2015XXVI}. We find no cross-matches, which confirms that the point-source masking strategy followed in our cluster finding pipeline successfully mitigates against the presence of bright point sources in the \textit{Planck} maps.

\subsection{Inspection of unconfirmed detections}\label{subsec:inspection}

With the goal of trying to shed light on the nature of the detections in the master catalogue without a positive cross-match with any of the external catalogues nor with the GCC catalogue (183 in total), we have inspected the MMF signal-to-noise maps at which each of them is detected. We have found two kinds of unconfirmed detections that we regard \emph{true} false detections and flagged them as such in the master catalogue by setting their confirmation flag to -1.

The first kind of true false detections is formed by detections made at the border of a given selection tile on a large angular scale, usually the largest one considered (32\,arcmin), and that are near a confirmed detection in the neighbouring tile but at an angular distance greater than 10\,arcmin. These detections can be explained as follows. Suppose there is a cluster near the edge of tile $B$, appearing therefore both in the field of tile $B$ and in that of its neighbouring tile, tile $A$. Suppose also that this cluster is detected in tile $B$. The cluster will probably also be detectable in the signal-to-noise maps of tile $A$ before the peak-finding mask of tile $A$ is applied in order to restrict peak finding to tile $A$. If the cluster is significant enough, the region around it above which the signal-to-noise is greater than the detection threshold in the field of tile $A$ may reach tile $A$. This is more likely to happen at a large search angular scale. In that case, the cluster will also be detected in tile $A$, typically at a large angular scale and with a poorly estimated sky position. We have identified 16 such detections in our master catalogue (with the iMMF6 catalogue having 5 of them).

The second kind of true false detection is analogous to the first kind, but applies to detections lying near the edge of a point-source-masked region, typically also found at large angular scales. These unconfirmed detections probably correspond to true objects lying inside the region masked out by the buffered point-source mask, being detected at large angular scales on the edge of the point-source masked region. We have found 25 such detections in our master catalogue (with the iMMF6 catalogue having 7 of them).

We have performed a similar inspection analysis to our validation catalogues, finding in them a comparable number of true false detections of the two kinds described here (see Section\,\ref{subec:purity_sim}).

In addition, we have found 2 non-cross-matched detections each of which is probably associated with another cross-matched detection separated from it by slightly more than the cross-matching distance of 10\,arcmin. As a consequence, we have flagged these detections as true false detections, setting their confirmation flag to -1. Finally, we have identified 2 selection tiles (number 436 and 437) featuring significant Fourier ringing that leads to a number of non-cross-matched detections (13 in total). These tiles lie near the Galactic plane and are significantly masked out by the Galactic mask, which may be responsible for the ringing. We have decided to also flag all these detections as true false detections, setting their confirmation flag to -1, and advise against the use of these two selection tiles in a cosmological analysis.

Taking all these 56 true false detections into account, we are left with a total of 127 non-cross-matched detections whose status as real clusters remains uncertain. We recall that there is a total of 1152 detections within the cosmology mask of which 882 have an available redshift measurement and 17 have been identified with a GCC.

\begin{figure*}
\centering
\includegraphics[width=0.9\textwidth,trim={0mm 0mm 0mm 0mm},clip]{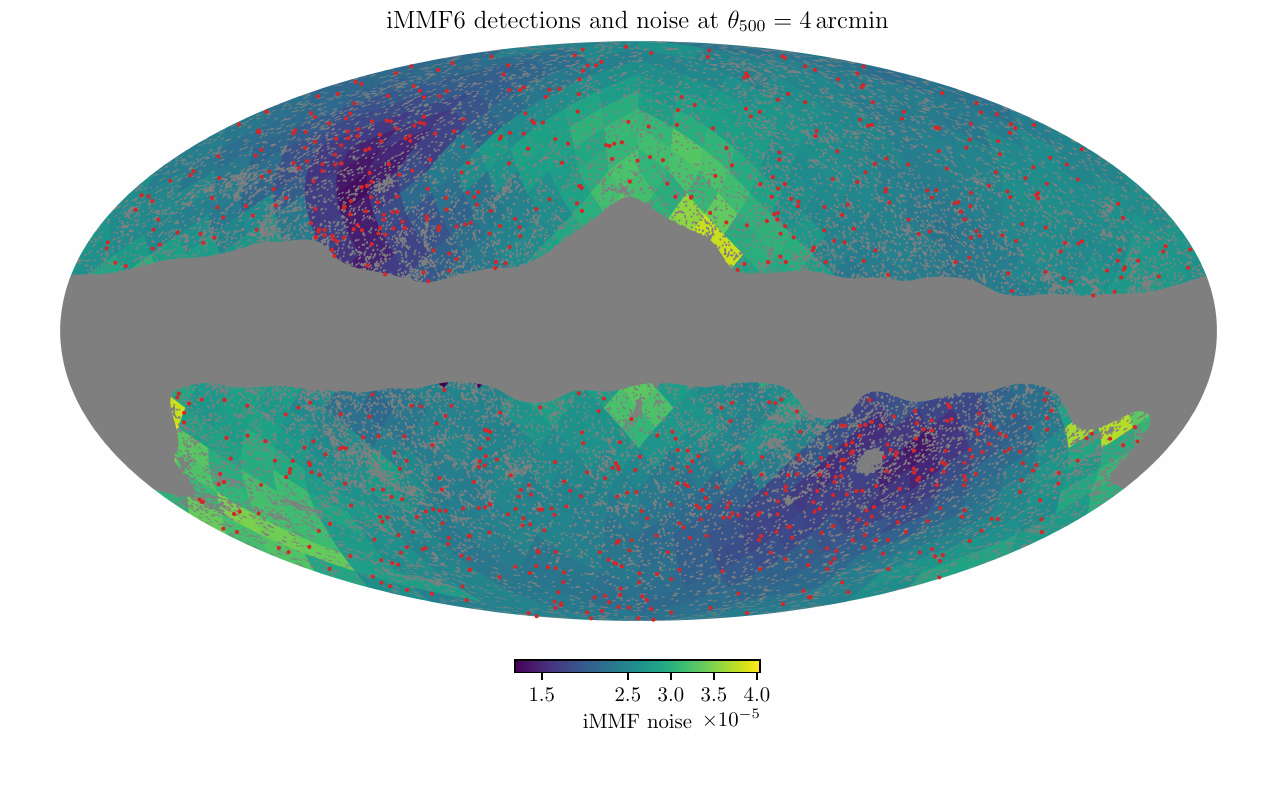}
\caption{Detections in the iMMF6 catalogue in Galactic coordinates (red points) shown over the multi-frequency matched filter (MMF) noise $\sigma_{\mathrm{f}}$ of each selection tile evaluated at $\theta_{500}=4$\,arcmin. The cosmology mask is shown as the grey region. We note that $\sigma_{\mathrm{f}}$ includes contributions both from instrumental noise and foregrounds (e.g., the CMB, the CIB, the kSZ effect and Galactic foregrounds).}
\label{fig:map_detections}
\end{figure*}

\section{The \textit{Planck} SZ\lowercase{i}F\lowercase{i} galaxy cluster catalogues: properties}\label{sec:catalogue}

In this section we introduce our new \textit{Planck} \texttt{SZiFi} catalogues, describing their properties. We recall that there are a total of 10 catalogues, each obtained with one of our 10 iterative MMFs (see Section\,\ref{subsec:matched_filters_used} and Table\,\ref{table:mmfs}), in addition to their 10 non-iterative counterparts. The catalogues are all compiled into a single master catalogue, as described in Section\,\ref{sec:confirmation}. The master catalogue will become publicly available at \href{https://github.com/inigozubeldia/szifi/tree/main/planck_szifi_master_catalogue}{this link} and a brief description of its entries can be found in Table\,\ref{table:catalogue}. 

Here, first, in Section\,\ref{subsec:sky}, we study the sky distribution of our detections. Next, in Section\,\ref{subsec:properties} we describe the distribution across both signal-to-noise and redshift and quantify how the signal-to-noise values of our detections change with the MMF employed. Then, in Section\,\ref{subsec:validation}, we study the confirmation and redshift assignment fraction and, finally, in Section\,\ref{subsec:impact_iterative} we analyse the impact of iterative noise covariance estimation.

Hereafter, when analysing the catalogues in the master catalogue, we will only consider the detections within the cosmology mask (see Section\,\ref{subsec:master_catalogue}) and with a confirmation flag of $0$ or $1$, discarding the detections that we have concluded to be false (see Section\,\ref{subsec:inspection}). We will also place a particular focus on the iterative catalogues, considering the non-iterative catalogues only when assessing the impact of iterative noise estimation and for comparison with the \textit{Planck} MMF3 catalogue, which was constructed without iterative noise covariance estimation (see Section\,\ref{sec:comparison}).

\subsection{Sky distribution}\label{subsec:sky}

\begin{figure}
\centering
\includegraphics[width=0.5\textwidth,trim={0mm 0mm 0mm 0mm},clip]{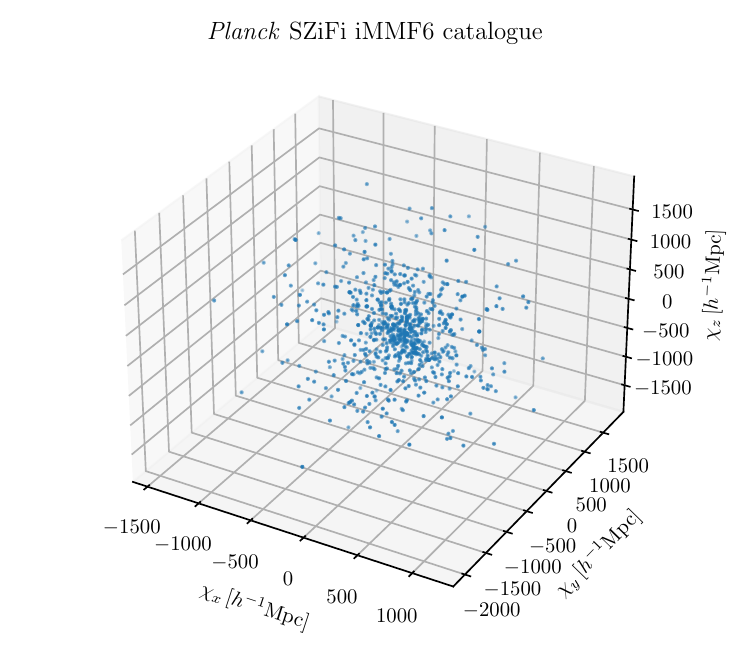}
\caption{Three-dimensional positions of the clusters in the iMMF6 catalogue with a redshift measurement in comoving coordinates. The orientation of the axes is given by the Galactic coordinate system, with the $z$ axis corresponding to a Galactic latitude of $90$\,deg and the $x$ axis to a Galactic longitude of $0$\,deg. The distances to each cluster have been computed assuming our reference cosmological model (see Section\,\ref{subsec:sky}).}
\label{fig:map_3d}
\end{figure}

Figure\,\ref{fig:map_detections} shows the sky position, in Galactic coordinates, of the detections in the iMMF6 catalogue (red points), our baseline catalogue, plotted over the MMF noise $\sigma_{\mathrm{f}}$ for each selection tile at $\theta_{500} = 4$\,arcmin. We recall that $\sigma_{\mathrm{f}}$ includes contributions due to both noise and foregrounds; see Section\,\ref{subsec:mmf} for how it is defined. The cosmology mask, which leaves a total of 58.65\,\% of the sky unmasked, is shown as the grey area. Its two components (Galactic and point-source) are clearly visible. The impact of the \textit{Planck} scanning strategy is also visible, with lower MMF noise $\sigma_{\mathrm{f}}$ and, therefore, more detections near the ecliptic poles. The other catalogues feature a similar sky distribution that is not shown here for conciseness.

In addition, Figure\,\ref{fig:map_3d} depicts the three-dimensional position, in comoving coordinates, of the subset of the detections in the iMMF6 catalogue for which a redshift measurement is available, all of which correspond to confirmed clusters. The orientation of the axes is given by the Galactic coordinate system, with the $z$ axis corresponding to a Galactic latitude of $90$\,deg and the $x$ axis to a Galactic longitude of $0$\,deg. The distances have been computed assuming a reference flat Lambda Cold Dark Matter ($\Lambda$CDM) cosmological model with $h=0.6774$, $\Omega_{\mathrm{c}} = 0.26603$, $\Omega_{\mathrm{b}} = 0.04897$, $A_s= 2.08467 \times 10^{-9}$, $n_s = 0.96$, and $\sum m_\nu = 0.06$\,eV, with the parameter values taken to be consistent with the \textit{Planck} CMB constraints \citep{Planck2018VI}. The region masked by the Galactic component of the mask can be clearly observed.

\subsection{Signal-to-noise and redshift distribution}\label{subsec:properties}

\begin{figure*}
\centering
\includegraphics[width=0.9\textwidth,trim={0mm 0mm 0mm 0mm},clip]{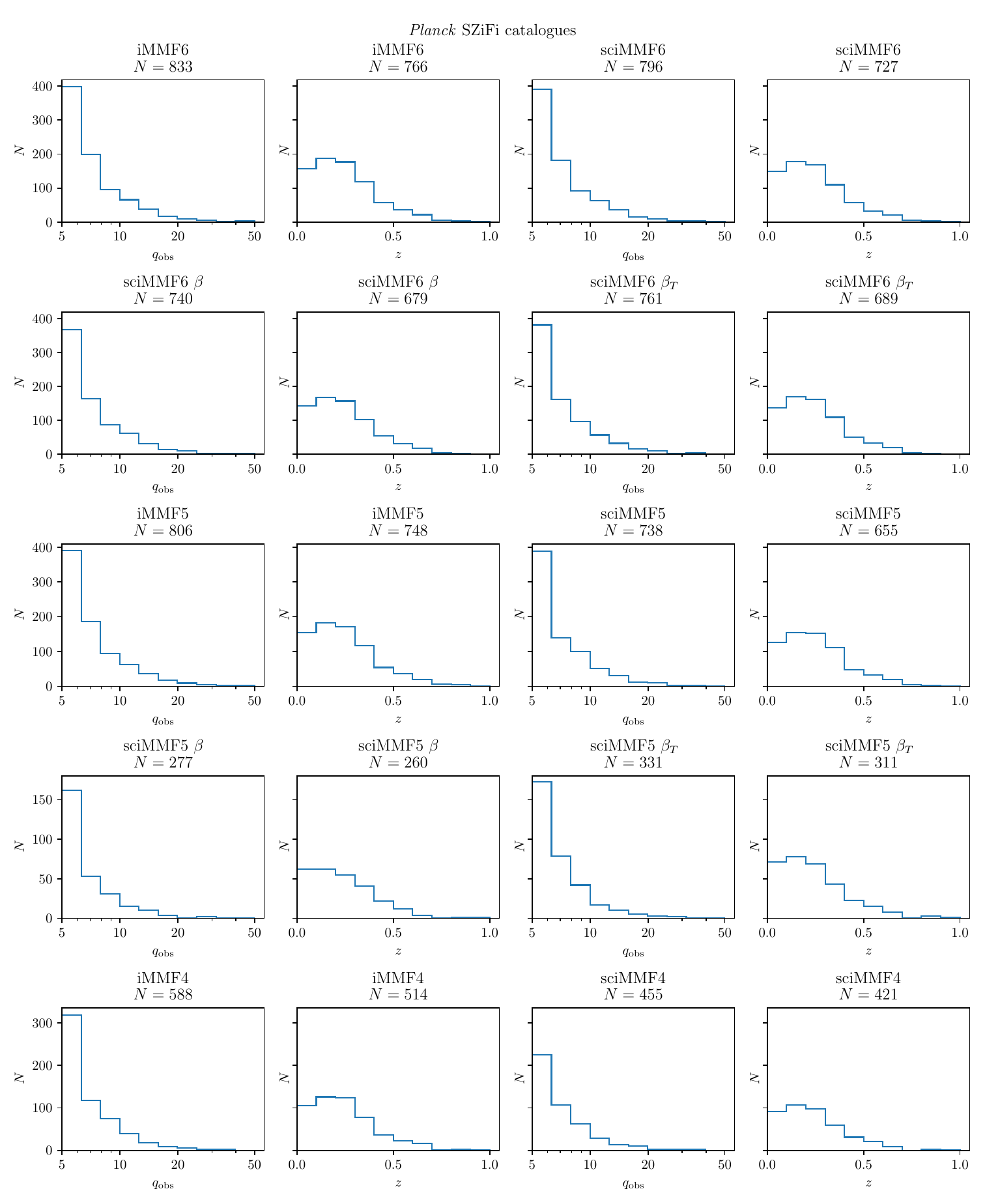}
\caption{Number counts in our 10 iterative cluster catalogues, across both signal-to-noise $q_{\mathrm{obs}}$ and redshift $z$ (left and right panel for each catalogue, respectively). The total number of detections in each catalogue is reported above the panel showing its corresponding counts across signal-to-noise, whereas the total number of confirmed detections with a redshift measurement is shown above the corresponding panel for the redshift counts. The iMMF6 and the sciMMF5 $\beta$ catalogues contain, respectively, the largest and smallest number of clusters.} 
\label{fig:histograms}
\end{figure*}

Figure\,\ref{fig:histograms} shows the number counts in the 10 iterative catalogues in the master catalogue, both across signal-to-noise $q_{\mathrm{obs}}$ (for all the detections in each catalogue) and across redshift (for the confirmed detections with an available redshift measurement). We have binned the detections across signal-to-noise in 10 bins logarithmically spaced between $q_{\mathrm{obs}} = 5$ and $q_{\mathrm{obs}} = 50$, and across redshift in 10 bins linearly spaced between $z=0$ and $z=1$. The total number of detections in each catalogue is reported above the panel showing its corresponding number counts across signal-to-noise, whereas the associated total number of confirmed detections with a redshift measurement is shown above its corresponding redshift counts panel.

As expected, the number counts decrease with signal-to-noise for every catalogue. This is also the case for the official \textit{Planck} catalogues \citep{{Planck2012I,Planck2013XXIX,Planck2016xxvii}}, as well as for other tSZ catalogues (e.g., \citealt{Hilton2018,Bleem2022}), and is a consequence of the steepness of the mass dependence of the halo mass function. The redshift distribution, on the other hand, peaks around $z \simeq 0.2$, as it does for the official \textit{Planck} catalogues \citep{{Planck2012I,Planck2013XXIX,Planck2016xxvii}}.

Spectral deprojection of the CIB and/or the use of fewer frequency channels results in fewer detections in every signal-to-noise or redshift bin. There are two contributions to this effect. First, there is a signal-to-noise penalty due to using fewer channels and/or spectral deprojection (see, e.g., \citealt{Zubeldia2023}). This penalty increases as fewer channels are used and/or more SEDs (including the moments) are deprojected. Second, the amplitude of potential biases in the signal-to-noise due to cluster-correlated emission, such as the CIB, may depend on the channel and/or deprojection combination (see below and also \citealt{Zubeldia2023}). As expected, the catalogue with the highest number of detections is the iMMF6 catalogue, which is constructed using all 6 HFI channels and without any spectral deprojection. It contains 767 detections, 694 of them with a redshift measurement. On the other hand, the catalogue with the lowest number of detections is the sciMMF5 $\beta$ catalogue, constructed using the 5 lowest-frequency HFI channels and spectrally deprojecting the CIB and its first-order moment with respect to $\beta$. It contains a total of 277 detections, 260 of them with a redshift measurement.

\begin{figure}
\centering
\includegraphics[width=0.5\textwidth,trim={0mm 0mm 0mm 0mm},clip]{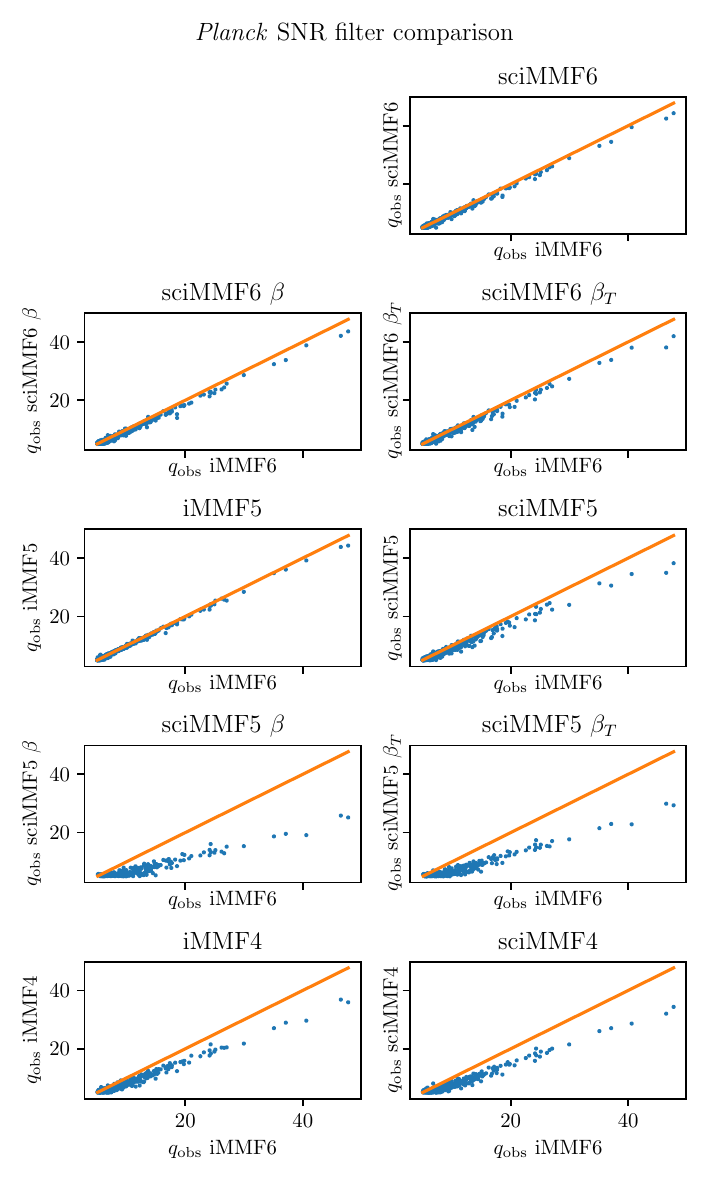}
\caption{Signal-to-noise values for the detections in each of our catalogues (identified by the title above each panel) plotted against the corresponding values in the iMMF6 catalogue for the common detections in each pair of catalogues. As expected, the iMMF6 catalogue leads to the highest signal-to-noise measurements, with the signal-to-noise decreasing with the use of fewer frequency channels and/or the application of more stringent spectral deprojection. The changes in the signal-to-noise seen here are responsible for the changes in the number counts observed in Figure\,\ref{fig:histograms}.}
\label{fig:snr_comparison}
\end{figure}

The change in the number counts when different MMFs are used can be understood as the result of the change in the signal-to-noise of each detection (or missed detection, if the signal-to-noise falls below the detection threshold for any given MMF). This is illustrated in Figure\,\ref{fig:snr_comparison}, which depicts the signal-to-noise for the detections in each catalogue (vertical axis) plotted against the corresponding signal-to-noise values in the iMMF6 catalogue (horizontal axis) for the common detections between each pair of catalogues. It is apparent that CIB spectral deprojection and/or the use of fewer channels reduces the signal-to-noise for any detection.

As mentioned above, there are two contributions to this signal-to-noise change: (i) the signal-to-noise penalty and (ii) different sensitivity to correlated emission, such as the CIB. These two different contributions are difficult to disentangle by considering the signal-to-noise measurements alone. We set out to do so by using the $\hat{y}_0$ MMF measurements instead, as we describe in detail in Sections\,\ref{sec:cib} and\,\ref{sec:rsz}.

We note that we do not provide any integrated Compton-$y$ measurements for the detections in our catalogues, regarding $\hat{y}_0$ and $q_{\mathrm{obs}}$ as the natural output quantities of our MMFs. However, in Section\,\ref{sec:cib} we use derived integrated Compton-$y$ measurements (specifically, integrated within $\theta_{500}$), obtained assuming the \citet{Arnaud2010} profile, in order to estimate the cluster electron temperatures. These derived measurements can be easily obtained from our catalogues using the scaling relations presented in Section\,\ref{sec:cib} (specifically, Eqs.\,\ref{eq:inty_angular} and\,\ref{eq:inty_physical}). In addition, in Section\,\ref{sec:validation}, we assess how these derived integrated measurements relate to their true values.

\subsection{Catalogue confirmation: purity and redshifts}\label{subsec:validation}

\begin{figure}
\centering
\includegraphics[width=0.35\textwidth,trim={0mm 0mm 0mm 0mm},clip]{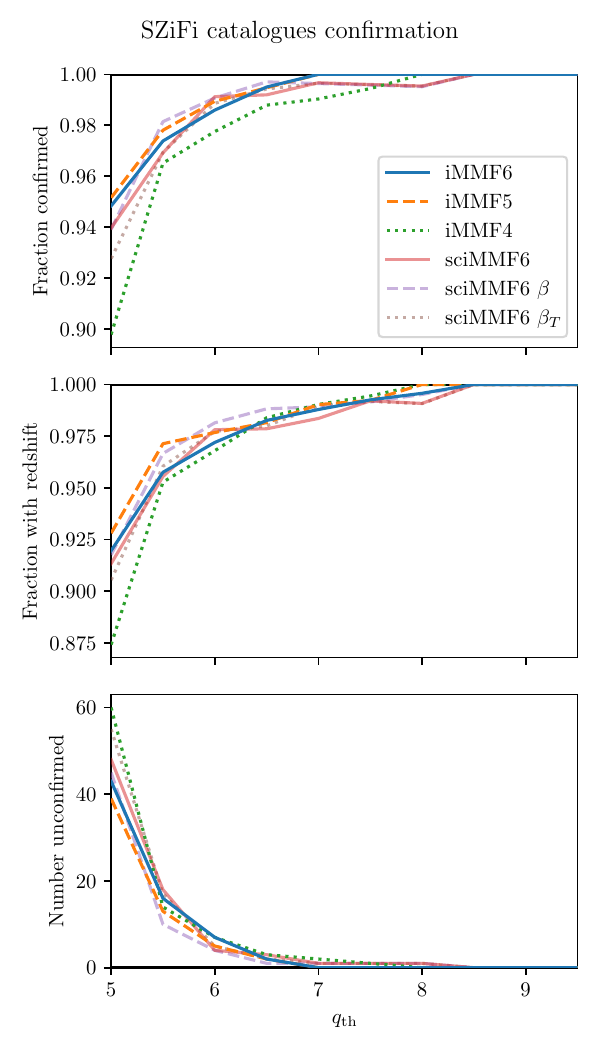}
\caption{Fraction of confirmed detections (upper panel) and of detections with a redshift measurement, all of which are confirmed clusters (middle panel), and number of unconfirmed detections (lower panel), as a function of minimum signal-to-noise $q_{\mathrm{th}}$, for six of our catalogues. Compare with the analogous results for the validation catalogues obtained ffrom our synthetic \textit{Planck}-like observations (Figure\,\ref{fig:validation_purity}).}
\label{fig:confirmation}
\end{figure}

The upper panel of Figure\,\ref{fig:confirmation} shows the fraction of detections in our catalogues that we have confirmed as real galaxy clusters in our cross-matching procedure (see Section\,\ref{sec:confirmation}) as a function of the minimum signal-to-noise $q_{\mathrm{th}}$ for six of our catalogues. This confirmation fraction can be understood as an empirical lower limit on the fraction of true detections in the catalogue, typically referred to as purity. As expected, the confirmation fraction increases with signal-to-noise. For most MMFs, including the baseline catalogue iMMF6, it goes up from about 0.95 at $q_{\mathrm{th}}=5$ to 1 at $q_{\mathrm{th}}=8$. The most notable exception is the iMMF4 catalogue, for which the confirmation fraction starts at a lower value of 0.9 at $q_{\mathrm{th}}=5$, but similarly rises to 1 at $q_{\mathrm{th}}=8$. Note, in particular, that spectral deprojection does not degrade the confirmation fraction significantly, with our sciMMFs having a confirmation fraction that is comparable to those for iMMF6 and iMMF5. The number of unconfirmed detections in each catalogue as a function of $q_{\mathrm{th}}$, i.e., the number of detections associated with each confirmation fraction, is shown in the lower panel of Figure\,\ref{fig:confirmation}.

The values of the confirmation fraction as a function of $q_{\mathrm{th}}$, as well as its behaviour across different MMFs, are broadly similar to what we observe for our validation catalogues, produced by applying our cluster finding pipeline to synthetic sky maps (see Section\,\ref{sec:validation}, and, in particular, Figure\,\ref{fig:validation_purity}). The associated numbers of unconfirmed detections are also similar. We note that this similarity should not be over interpreted, as we do not claim our synthetic data to be a perfect description of the \textit{Planck} sky, particularly regarding, e.g., point sources that may be spuriously detected by our pipeline.

Finally, the middle panel of Figure\,\ref{fig:confirmation} shows the fraction of detections to which we have been able to assign a redshift measurement, also as a function of minimum signal-to-noise $q_{\mathrm{th}}$. As with the confirmation fraction, this fraction increases with signal-to-noise, rising from about 0.92 at $q_{\mathrm{th}}=5$ for all catalogues except iMMF4 to 1 at $q_{\mathrm{th}}=8.5$ for all the catalogues.

\subsection{Impact of iterative noise covariance estimation}\label{subsec:impact_iterative}

\begin{figure}
\centering
\includegraphics[width=0.4\textwidth,trim={0mm 0mm 0mm 0mm},clip]{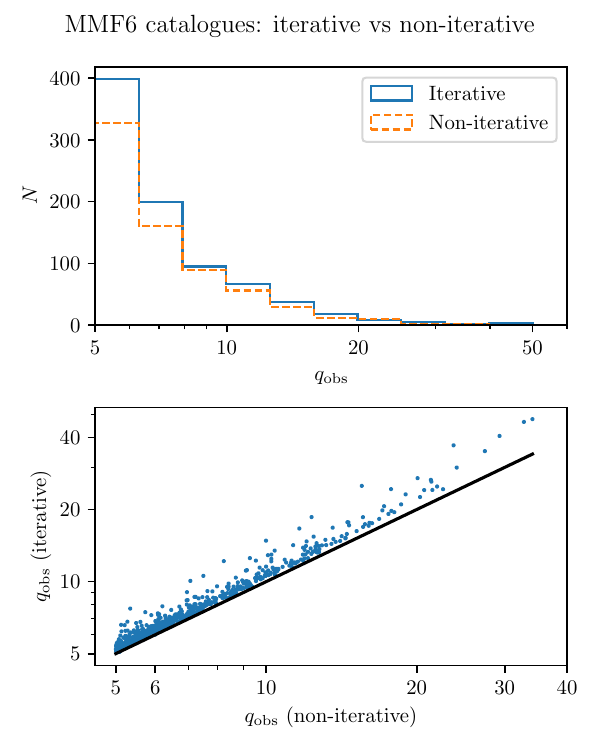}
\caption{\textit{Upper panel}: Number counts in our iMMF6 and MMF6 catalogues (solid blue and dashed orange histograms, respectively) across signal-to-noise $q_{\mathrm{obs}}$. We recall that the iMMF6 catalogue is the catalogue obtained with all 6 HFI channels applying iterative noise covariance estimation, whereas the MMF6 catalogue is its non-iterative counterpart. \textit{Lower panel}: Signal-to-noise measurements in the iMMF6 catalogue (vertical axis), plotted against the corresponding values in the MMF6 catalogue (horizontal axis) for the common detections. As expected, iterative noise covariance estimation results in an increase in the signal-to-noise of each detection due to the correct estimation of the noise covariance and to the removal of the negative covariance bias. This, in turn, leads to a larger number of objects in each signal-to-noise bin as seen in the upper panel.}
\label{fig:comparison_it_noit}
\end{figure}

With the goal of investigating the impact of iterative noise covariance estimation, we show the number counts in the iMMF6 and MMF6 catalogues as a function of signal-to-noise in the upper panel of Figure\,\ref{fig:comparison_it_noit}. We recall that the only difference between these two catalogues is that the iMMF6 catalogue is constructed with iterative noise covariance estimation, MM6 being its non-iterative counterpart. As expected, iterative noise covariance estimation increases the number counts in every signal-to-noise bin. This effect can be understood as a positive change in the signal-to-noise of each detection. This change can be clearly seen in the lower panel of Figure\,\ref{fig:comparison_it_noit}, which shows the signal-to-noise values for the iMMF6 catalogue plotted against those in the MMF6 catalogue, for the common detections in both catalogues. There it is apparent that iterative noise covariance estimation increases the signal-to-noise of all detections, with the magnitude of the boost increasing with signal-to-noise. These results are in agreement with the findings for synthetic observations reported in \citet{Zubeldia2022} (see, in particular, their Figures\,6 and\,7).

There are two contributions to this increase in the signal-to-noise. First, iterative noise covariance estimation prevents the overestimation of noise covariance due to the presence of the signal, thereby boosting the signal-to-noise. Moreover, it eliminates the negative covariance bias, further enhancing the signal-to-noise (see Section\,\ref{subsec:iterative} and \citealt{Zubeldia2022}).

\section{Comparison with the \textit{Planck} MMF3 catalogue}\label{sec:comparison}

\begin{figure*}
\centering
\includegraphics[width=0.9\textwidth,trim={0mm 0mm 0mm 0mm},clip]{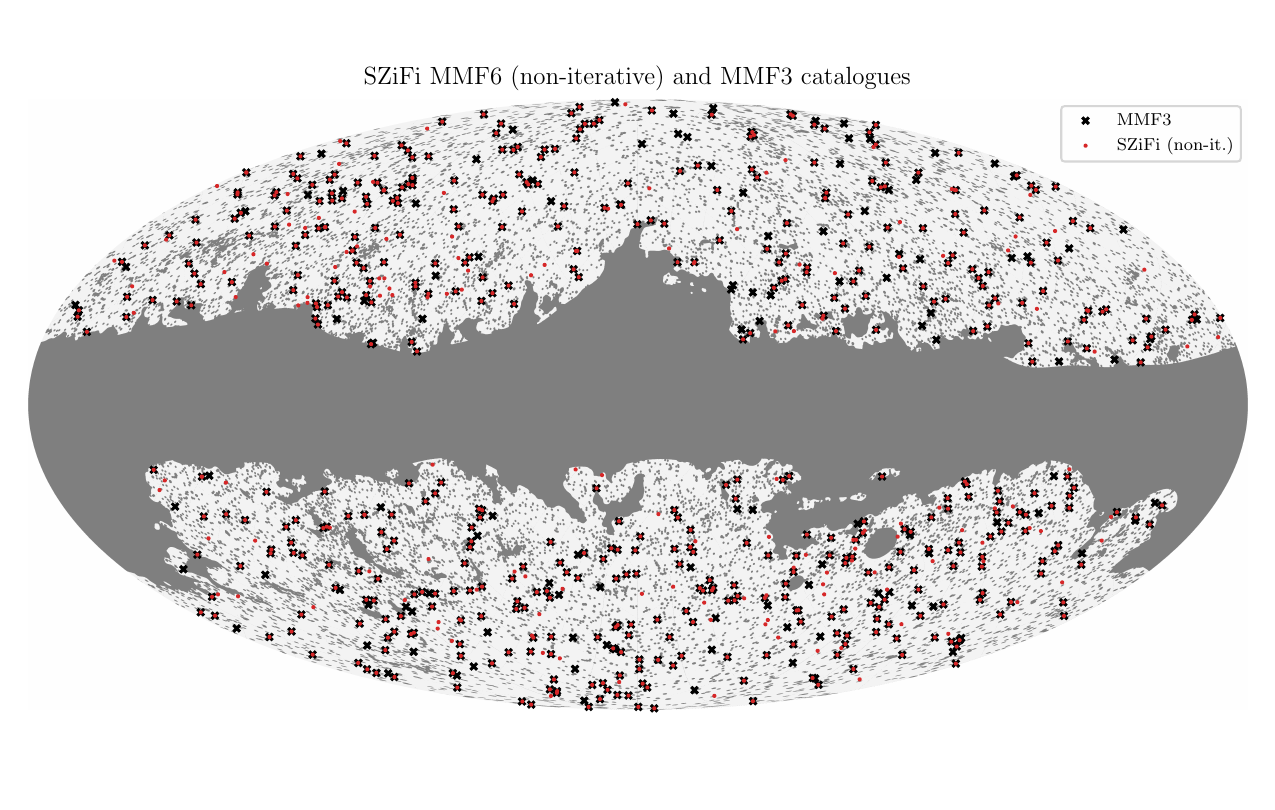}
\caption{Detections in our MMF6 catalogue (red points) and in the latest official \textit{Planck} catalogue, the MMF3 catalogue (black crosses), in Galactic coordinates. The union of our cosmology mask and of the \textit{Planck} PSZ2 survey mask (the `common mask') is shown as the shaded region.}
\label{fig:map_detections_comparison}
\end{figure*}

\begin{figure}
\centering
\includegraphics[width=0.5\textwidth,trim={0mm 0mm 0mm 0mm},clip]{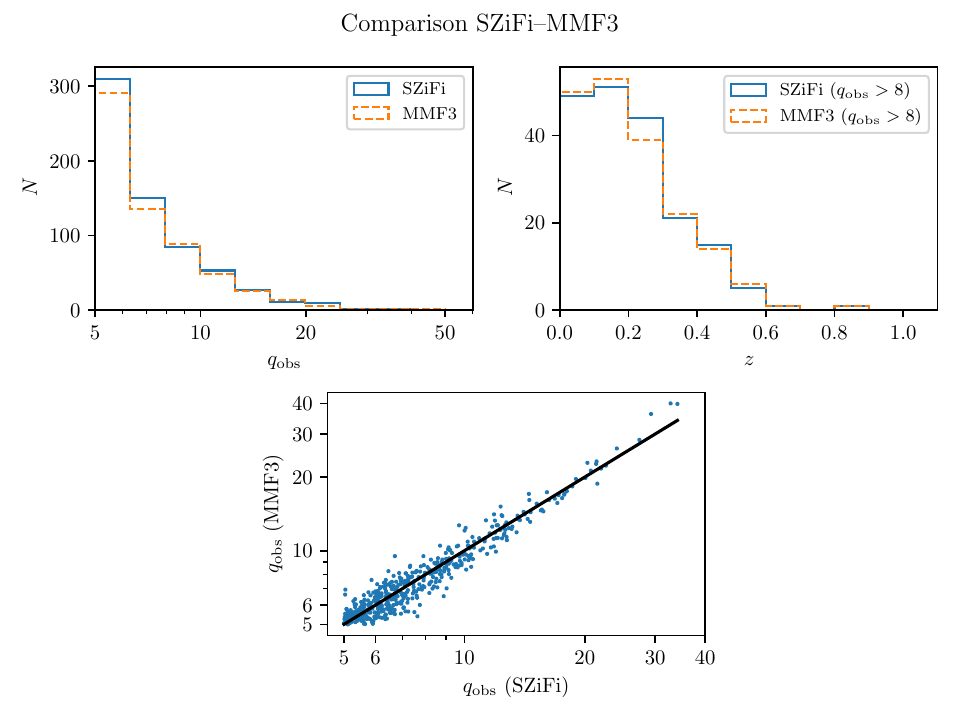}
\caption{Number counts in our \texttt{SZiFi} MMF6 catalogue and in the official \textit{Planck} MMF3 catalogue (solid blue and dashed orange histograms, respectively) binned as a function of both signal-to-noise (upper left panel) and redshift (upper right panel), for the common mask. In the upper right panel, only the clusters with $q_{\mathrm{obs}} > 8$ are shown, to ensure that all the clusters in the sample have an available redshift measurement. The lower panel shows the signal-to-noise values in the  \texttt{SZiFi} MMF6 and the MMF3 catalogues for the common detections in both catalogues.}
\label{fig:comparison_szifi_mmf3}
\end{figure}

In this section we compare our MMF6 catalogue with the \textit{Planck} cluster catalogue that was used in the latest official \textit{Planck} cosmological analysis, the MMF3 catalogue, which forms part of the \textit{Planck} PSZ2 catalogue \citep{Planck2016xxvii}. 

The MMF3 catalogue was obtained using a similar detection pipeline to the one employed in this work. The sky was tessellated into tiles and a matched filter was applied to each tile for a set of angular scales (see \citealt{Planck2016xxvii}). The same cluster profile was used for the matched filter template, the same range of angular scales was considered, and a standard MMF was employed, using the 6 HFI channels, as in our MMF6 catalogue. However, the MMF3 catalogue used the \textit{Planck} PR2 HFI frequency maps instead of those from PR3 used in this work. In addition, a different tessellation scheme, a less aggressive point-source mask, and a slightly different Galactic mask were applied, and a different multipole range was used in the matched filtering process (all the modes in the cut-outs, instead of $l_{\mathrm{min}} = 100$ and $l_{\mathrm{max}} = 2500$ as used here). The noise covariance was estimated non-iteratively assuming it to be equal to the covariance of the data. Despite these differences, the MMF3 catalogue can therefore be most suitably compared with our MMF6 catalogue, as, we recall, the MMF6 catalogue was also obtained using a standard MMF for all 6 HFI channels and without iterative noise covariance estimation.

Figure\,\ref{fig:map_detections_comparison} shows the sky position of the detections in our MMF6 catalogue (red points) and in the MMF3 catalogue (black crosses) in Galactic coordinates. The union of our cosmology mask and of the PSZ2 survey mask (the `common mask') is shown as the grey region. Only the detections falling outside of this common mask are shown; we note that in the rest of this section we will compare the MMF6 and MMF3 catalogues restricted to this sky region. As it can be seen in Figure\,\ref{fig:map_detections_comparison}, there are many common detections between the two catalogues, as well as some detections present only in either catalogue.

Further comparison between the MMF6 and MMF3 catalogues can be achieved by examining the number counts across signal-to-noise and redshift for the common mask, as well as the signal-to-noise values for the common detections. This comparison is illustrated in Figure\,\ref{fig:comparison_szifi_mmf3}. The upper left panel shows the number counts of detections in both catalogues (including unconfirmed detections) for the common mask (solid blue histogram for MMF6, dashed orange histogram for MMF3). The counts are similar, with the MMF6 catalogue having slightly more detections in the first two signal-to-noise bins. The lower panel shows the signal-to-noise values for the common detections in both catalogues (MMF6 on the horizontal axis and MMF3 on the vertical axis). There is very good agreement between the two catalogues, with the only notable difference being slightly larger signal-to-noise values for MMF3 at the high signal-to-noise end. As expected, there is some scatter. This scatter can be attributed to differences in the maps used (PR2 vs PR3) and in the detection pipeline, such as the tessellation scheme, the bins in Fourier space used for noise covariance estimation, and the multipole range used for matched filtering. This scatter explains why, within the common mask, the MMF6 catalogue includes some detections missing in the MMF3 catalogue and vice versa: near the detection threshold, some objects fluctuate above it in one catalogue and below it in the other one.

Regarding the redshift distribution, the upper right panel of Figure\,\ref{fig:comparison_szifi_mmf3} shows the number counts in the MMF6 and MMF3 catalogues (solid blue and dashed orange histograms, respectively) across redshift for detections above a signal-to-noise threshold of $q_{\mathrm{th}} = 8$. We have imposed $q_{\mathrm{th}} = 8$ to ensure that all the detections in both catalogues have a redshift measurement (see Figure\,\ref{fig:confirmation}). The redshift distributions for MMF6 and MMF3 are consistent with each other, with both peaking in the $0.1$--$0.2$ redshift bin.



\section{Impact of the CIB}\label{sec:cib}

In this section we investigate the impact of the cluster-correlated CIB on our cluster catalogues. We first explain how we set out to do so in Section\,\ref{subsec:reextract}. Next, we describe the synthetic observations that we use for comparison with our real measurements in Section\,\ref{subsec:sims_cib}, and we discuss our findings in Section\,\ref{subsec:cib_results}.

\subsection{Re-extracting $\hat{y}_0$}\label{subsec:reextract}

As discussed in Section\,\ref{subsec:matched_filters_used}, any signal that is spatially correlated with the cluster tSZ signal, such as the CIB, can bias the inferred values of the cluster observables. This bias may, in turn, affect number-count cosmological constraints (see also \citealt{Melin2018,Zubeldia2023}). Here, we assess the magnitude of the bias caused by the cluster-correlated CIB (the `CIB bias') by considering the MMF amplitude parameter $\hat{y}_0$ instead of our preferred cluster observable, the signal-to-noise $q_{\mathrm{obs}}$.

The reason for this approach is that $q_{\mathrm{obs}}$ suffers from a signal-to-noise penalty when fewer frequency channels are used and/or when the CIB is spectrally deprojected, as discussed in Section\,\ref{subsec:properties}. These two effects are difficult to disentangle when considering the signal-to-noise alone. Ignoring selection effects, the value of $\hat{y}_0$, however, is not affected by the signal-to-noise penalty: only its measurement uncertainty is. This property makes $\hat{y}_0$ a more suitable observable for investigating the impact of the CIB bias on our cluster catalogues.

The $\hat{y}_0$ estimate delivered by the cluster detection pipeline for each cluster is, however, expected to be biased. Indeed, as noted in Section\,\ref{subsec:peak}, $\hat{y}_0$ can be thought of a maximum-likelihood estimate. Maximum-likelihood estimates are, in general, biased (see, e.g., \citealt{Zubeldia2021}, and also Section\,\ref{subsec:peak}); for $\hat{y}_0$, this maximum-likelihood bias is analogous to the optimisation bias that affects the signal-to-noise (see Sections\,\ref{subsec:peak} and\,\ref{sec:validation} for brief discussions on the optimisation bias). In addition, $\hat{y}_0$ is expected to suffer from a Malmquist-type bias, since the sample is selected on the signal-to-noise, which is strongly correlated with $\hat{y}_0$. When comparing $\hat{y}_0$ values from two different MMFs for a given cluster detected with both MMFs, this bias is expected to be stronger for the MMF featuring a lower signal-to-noise (e.g., due to using fewer channels), as the cluster will lie closer to the selection threshold for that MMF than for the other one.

Due to the presence of these biases in the `blind' $\hat{y}_0$ measurements, we decide to consider only the clusters detected across all 10 iterative MMFs. We refer to this subsample of the master catalogue, containing a total of 247 clusters, as the `CIB subsample'. For the detections in this CIB subsample, we re-extract $\hat{y}_0$ at the average sky position across the 10 MMFs and at a fixed angular scale of $\theta_{500} = 5\,$arcmin. This re-extraction is performed using \texttt{SZiFi}'s fixed mode with iterative noise covariance estimation and all other pipeline specifications set to match those used for cluster detection (see Section\,\ref{subsec:pipeline}). The re-extraction is expected to reduce both the maximum-likelihood bias and the Malmquist-type bias.

However, these `fixed' $y_0$ measurements will still be biased due to the mismatch between the angular scale and sky position used for MMF extraction and their true values, an effect commonly known as miscentering for the sky position. In addition, any significant differences between the cluster profile assumed in the MMF and the true cluster profile will also introduce a bias. It is reasonable to assume, however, that these biases will be consistent across all MMFs and that the differences in the $\hat{y}_0$ measurements across the various MMFs will be dominated by correlated emission, with a potential additional contribution from the Malmquist-type bias and other potential selection effects, which we will quantify with synthetic observations. These differences across the different MMFs will be the focus of the remainder of this section.

\subsection{Synthetic observations}\label{subsec:sims_cib}

We apply our cluster detection and $\hat{y}_0$ re-extraction pipelines to two sets of synthetic \textit{Planck}-like maps: one in which the CIB is spatially correlated with the tSZ signal and another one in which this correlation is artificially removed. We do this for two reasons. First, the synthetic observations in which the CIB is not correlated with the tSZ signal provide a way of quantifying the magnitude of the selection biases in our fixed $\hat{y}_0$ measurements. Second, the synthetic observations for which the CIB is correlated with the tSZ signal provide a prediction for the magnitude of the CIB-induced bias on the fixed $\hat{y}_0$ measurements that can then be contrasted with the real fixed $\hat{y}_0$ measurements.

We use the same synthetic maps that were used to assess CIB contamination of MMF cluster detection in \citet{Zubeldia2023}. These maps consist of full-sky temperature maps at the six frequency channels of the \textit{Planck} HFI. They are pixelised using the HEALPix pixelisation scheme \citep{Gorski2005} with $N_{\mathrm{side}} = 2048$ that used in the real \textit{Planck} HFI maps. Each temperature map contains the following components, all of which are convolved with the corresponding \textit{Planck} RIMO beam and the HEALPix pixel transfer function (except the instrumental noise, which is only convolved with the pixel transfer function):
\begin{itemize}
    \item \textbf{tSZ:} The Compton-$y$ map is taken from the Websky simulation and is rescaled at each frequency by the non-relativistic tSZ SED. Websky\footnote{\texttt{mocks.cita.utoronto.ca/websky}} \citep{Stein2019,Stein2020} is an all-sky second-order Lagrangian perturbation theory lightcone simulation with a minimum halo mass $M_{200} \approx 1.4 \times 10^{12} M_{\odot}$. Since the Websky Compton-$y$ map has $N_{\mathrm{side}} = 4096$, we degrade it to $N_{\mathrm{side}} = 2048$ with the HEALPix \texttt{ud\_grade} function after beam convolution. We degrade in a similar way all the Websky maps that we employ (kSZ, lensed CMB, and CIB). We neglect the effect of relativistic temperature corrections to the tSZ signal \citep[e.g.,][]{Itoh1998, Challinor1998, Sazonov1998, Chluba2012SZpack}, which we will quantify with the real \textit{Planck} data in Section\,\ref{sec:rsz}.
    \item \textbf{kSZ:} The kinetic SZ (kSZ) signal is also taken from the Websky simulation and includes both the late-time and the reionisation contributions \citep{Stein2020}.
    \item \textbf{Lensed CMB:} The lensed CMB is also taken from the Websky simulation.
    \item \textbf{CIB:} The CIB is also taken from the Websky simulation, which provides a map for each of the \textit{Planck} HFI frequencies. At the \textit{Planck} frequencies, the Websky CIB emission from redshift $z$ is given by a modified blackbody (see Eq.\,\ref{eq:sed}), with emissivity index $\beta = 1.6$, dust temperature at $z=0$ $T_0 = 20.7$\,K, and $\alpha=0.2$ \citep{Stein2020}.
    \item \textbf{Instrumental noise:} We generate a full-sky realisation of Gaussian white noise with noise levels, for increasing frequencies, of 77.4\,$\mu$K\,arcmin, 33\,$\mu$K\,arcmin, 46.8\,$\mu$K\,arcmin, 153.6\,$\mu$K\,arcmin, 46.8\,kJy\,sr$^{-1}$\,arcmin, and 43.2\,kJy\,sr$^{-1}$\,arcmin \citep{Planck2016VIII}. The noise is uncorrelated between different channels, and, as noted above, is convolved only by the HEALPix pixel transfer function, and not by the instrument beam.  
\end{itemize}

As in \citet{Zubeldia2023}, our maps do not contain Galactic foregrounds and radio point sources. However, the Websky CIB includes bright point-like sources that can significantly contaminate cluster detection. In order to minimise their impact, we construct a point-source mask by running \texttt{SZiFi} as a point-source finder on each of the six synthetic frequency maps, producing a point-source catalogue for each frequency map. \texttt{SZiFi} is applied using iterative noise covariance estimation, and a signal-to-noise selection threshold of 5 is imposed. We then merge the individual frequency catalogues to produce a final point-source catalogue. Finally, a binary point-source mask is constructed by zeroing all the pixels within a masking distance of each detected point source. We use twice the beam FWHM of the channel in which each point source was detected as the masking angular distance.

We apply our \texttt{SZiFi}-based cluster detection pipeline to these synthetic frequency maps, using our \texttt{SZiFi} point-source mask and the \textit{Planck} PR2 80\% Galactic mask that we also used for cluster detection in the real \textit{Planck} maps. We use the same \texttt{SZiFi} configuration (multipole ranges, matched filter template, etc.) as in our analysis of the real \textit{Planck} maps (see Section\,\ref{subsec:pipeline}) and produce catalogues for our 10 iterative MMFs (see Section\,\ref{subsec:matched_filters_used}). We then re-extract $\hat{y}_0$ for the common detections across all 10 MMFs at the average MMF sky position and at $\theta_{500} = 5$\,arcmin, as we did for the real \textit{Planck} detections in the CIB subsample.

We perform this cluster detection and $\hat{y}_0$ re-extraction procedure twice, producing two sets of catalogues. In the first instance, the Websky CIB is left as it is, spatially correlated with the tSZ signal, producing our `correlated Websky catalogues'. In the second instance, we artificially remove this correlation, producing our `randomised Websky catalogues'. Since \texttt{SZiFi} processes each selection tile independently, we remove the correlation by assigning to each tile the CIB cut-out corresponding to another randomly chosen tile, instead of the CIB cut-out associated with it. For reference, there are a total of 290 common detections across the correlated Websky catalogues and a total of 428 across the randomised ones.

We note that, for our synthetic simulations, the CIB is the only component that can potentially bias our cluster tSZ measurements. Indeed, although the Websky kSZ field is spatially correlated with the tSZ one, it averages out for our clusters due to being proportional to the cluster velocity along the line of sight. This was numerically verified in \citet{Zubeldia2022, Zubeldia2023}, where it was shown that, for Websky, removing the CIB correlation leads to unbiased cluster tSZ observables.

\subsection{Results and discussion}\label{subsec:cib_results}

\begin{figure}
\centering
\includegraphics[width=0.5\textwidth,trim={0mm 0mm 0mm 0mm},clip]{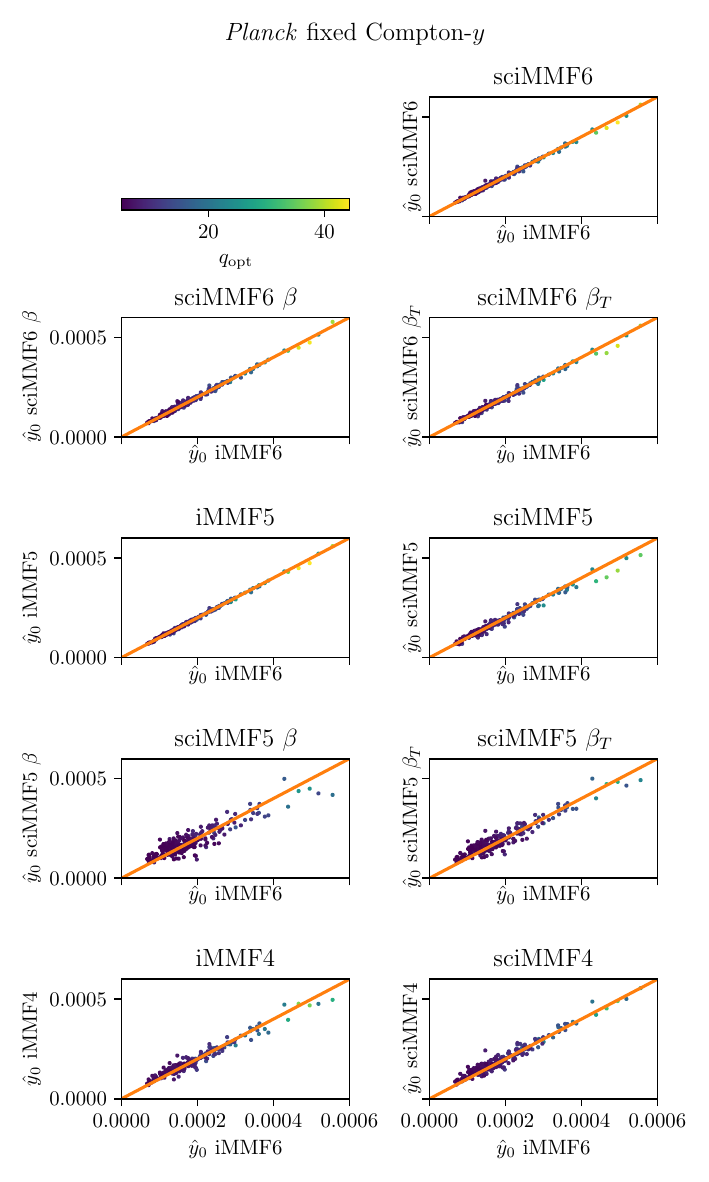}
\caption{Measurements of the MMF amplitude parameter $\hat{y}_0$ extracted at the average sky position across all 10 iterative MMFs and at a fixed angular scale $\theta_{500}=5$\,arcmin for the common detections in the 10 MMFs (247 detections). In each panel, the vertical axis corresponds to the panel's MMF (indicated in the panel title), whereas the horizontal axis corresponds to iMMF6. The data points are colour-coded by their detection signal-to-noise in the panel's MMF. It is apparent that CIB deprojection and/or the use of fewer frequency channels has little impact on $\hat{y}_0$  (see Section\,\ref{subsec:cib_results}).}
\label{fig:fixed_data}
\end{figure}


Figure\,\ref{fig:fixed_data} shows the re-extracted fixed $\hat{y}_0$ measurements in the CIB subsample for the real data. In each panel, the vertical axis corresponds to $\hat{y}_0$ for the panel's MMF (indicated in the panel title), while the horizontal axis corresponds to the same quantity for iMMF6. The data points are colour-coded by their detection signal-to-noise in the panel's MMF. As seen in Figure\,\ref{fig:fixed_data}, CIB deprojection and/or the use of fewer frequency channels have little impact on $\hat{y}_0$. A small shift is observed for sciMMF5 and sciMMF5 $\beta$, and perhaps also for sciMMF5 $\beta_T$ and iMMF4, at large $\hat{y}_0$ values. These shifts, however, are of low significance due to the small number of clusters at the high $\hat{y}_0$ end of the distribution.

These findings imply that CIB contamination of the $\hat{y}_0$ and signal-to-noise $q_{\mathrm{obs}}$ measurements in our catalogues is expected to be minimal. In particular, they imply that the catalogue featuring the largest overall signal-to-noise, the iMMF6 catalogue, contains little CIB contamination. Therefore, it may be possible to use it for cosmological inference without modelling this contamination, although further careful assessment at the likelihood analysis level is required.

\begin{figure}
\centering
\includegraphics[width=0.5\textwidth,trim={0mm 0mm 0mm 0mm},clip]{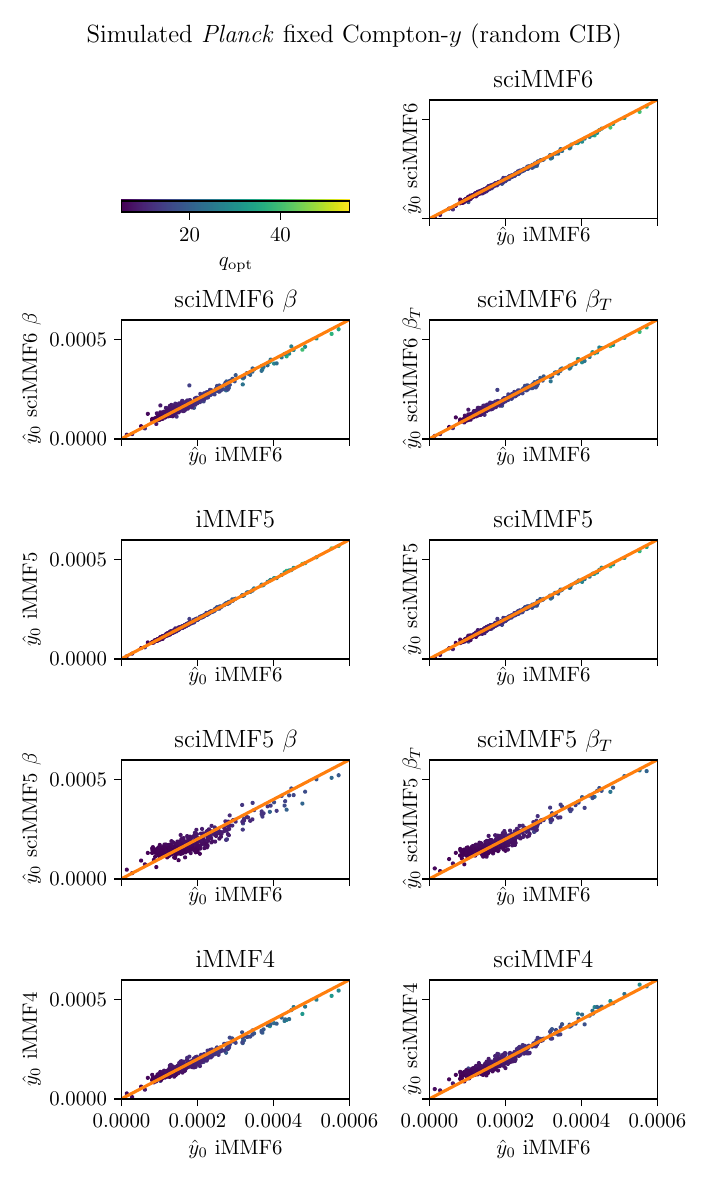}
\caption{As Figure\,\ref{fig:fixed_data}, but for our randomised Websky catalogues, which are obtained by applying the cluster detection and $\hat{y}_0$ re-extraction pipelines to our \textit{Planck}-like Websky-based synthetic maps in which the spatial correlation between the CIB and the tSZ signal has been artificially removed.}
\label{fig:fixed_sim_random}
\end{figure}

\begin{figure}
\centering
\includegraphics[width=0.5\textwidth,trim={0mm 0mm 0mm 0mm},clip]{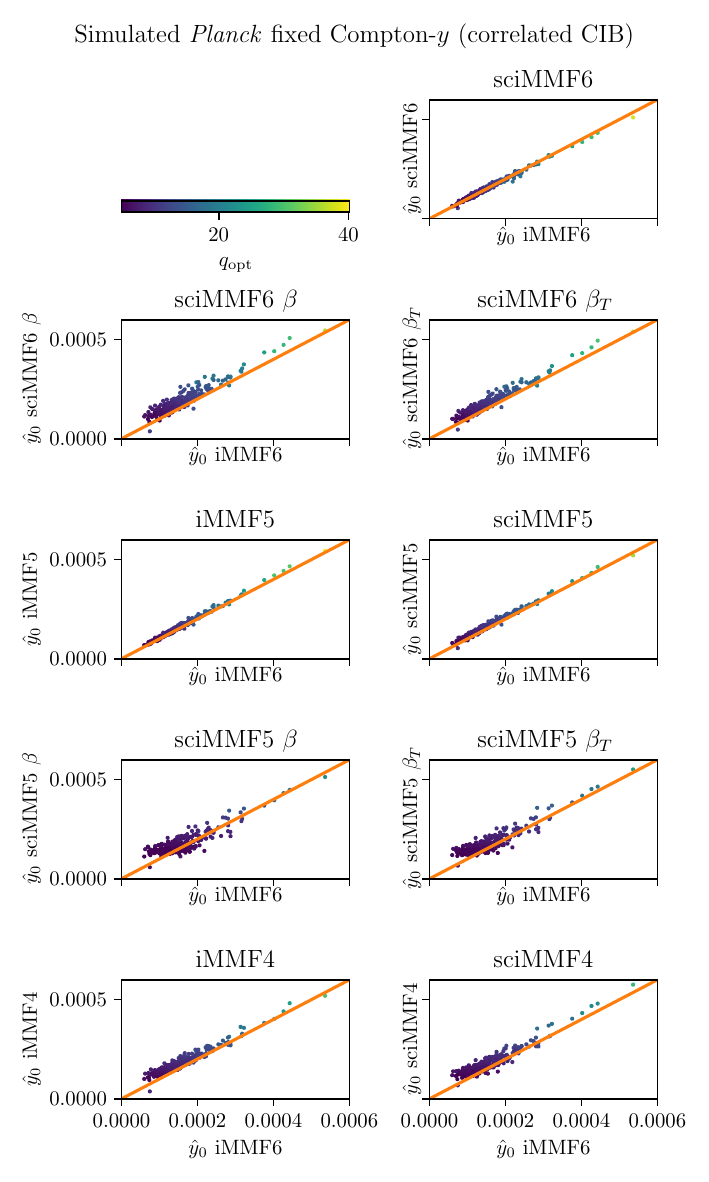}
\caption{As Figure\,\ref{fig:fixed_sim_random}, but with the spatial correlation between the CIB and tSZ signals in the synthetic maps, both of which come form the Websky simulation, left untouched.}
\label{fig:fixed_sim_correlated}
\end{figure}

Figures\,\ref{fig:fixed_sim_random} and\,\ref{fig:fixed_sim_correlated} are analogous to Figure\,\ref{fig:fixed_data}, but show instead the re-extracted fixed measurements for the randomised and correlated synthetic Websky catalogues, respectively. The distributions of the data points for the randomised catalogues are similar to those for the real \textit{Planck} data seen in Figure\,\ref{fig:fixed_data}. At large $\hat{y}_0$ values, most MMFs produce $\hat{y}_0$ measurements that agree with the iMMF6 ones. Small shifts are observed, as for the real data, for sciMMF5 and sciMMF5 $\beta$, and perhaps also for sciMMF5 $\beta_T$ and iMMF4.

We attribute the shifts at the large $\hat{y}_0$ end for the randomised sciMMF5, sciMMF5 $\beta$, sciMMF5 $\beta_T$, and iMMF4 catalogues to a selection effect, as they are also obtained from synthetic observations in which the tSZ signal is not correlated with any other signals. Indeed, even though these $\hat{y}_0$ measurements are not the blind, maximum-likelihood ones, they are extracted from the same maps. Therefore, their fluctuations are also expected to be correlated with the detection signal-to-noise and, as a consequence, they are also expected to be affected by the sample selection. Notably, very similar shifts are observed for the real catalogues obtained with the same MMFs  (see Figure\,\ref{fig:fixed_data}). This indicates that the shifts seen in these real catalogues can be attributed, at least partially, to this selection effect. 

A selection effect may also explain the shapes of the data point distribution at the low $\hat{y}_0$ end in the two last rows of panels in Figure\,\ref{fig:fixed_sim_random}, with larger $\hat{y}_0$ measurements for those MMFs than for iMMF6. Note that similar shapes are also observed for the real measurements in Figure\,\ref{fig:fixed_data}. As the low  $\hat{y}_0$ end corresponds to objects lying very close to the detection threshold (see the colour code), we attribute these shapes to the Malmquist-type bias mentioned in Section\,\ref{subsec:reextract}: for our CIB subsample, those MMFs result in significantly lower signal-to-noise values than iMMF6 does, and hence their $\hat{y}_0$ measurements are more strongly Malmquist-biased, i.e., larger.

To further investigate the origin of these features in the randomised case (shifts at large $\hat{y}_0$ values, distribution shapes at low $\hat{y}_0$ values), we extract $\hat{y}_0$ from our Websky synthetic maps for a mass-selected cluster sample. In this case, both features disappear, allowing us to fully attribute them to the sample selection. The results for our mass-selected sample are discussed in more detail in Appendix\,\ref{appendix:mass_selected}.

On the other hand, as seen in Figure\,\ref{fig:fixed_sim_correlated}, when the Websky CIB is spatially correlated with the tSZ signal there are significant shifts in the extracted $\hat{y}_0$ values compared to the randomised case. These shifts result from the spatial correlation between the Websky CIB and tSZ fields, which biases $\hat{y}_0$ for iMMF6. This bias is reduced or completely removed by using fewer frequency channels and/or spectral deprojection. Indeed, in \citet{Zubeldia2023} it was shown, using the same \textit{Planck}-like synthetic observations, that spectrally constrained MMFs are highly effective at suppressing the CIB bias, especially when the first-order moment with respect to either $\beta$ or $\beta_T$ is deprojected, delivering CIB-free signal-to-noise and Compton-$y$ measurements (see their Figures\,5, 9, and\,A2). It must be noted, however, that the significant difference between the shifts in Figure\,\ref{fig:fixed_sim_correlated} and those observed in the real measurements (Figure\,\ref{fig:fixed_data}) indicates that the Websky CIB model around the large-mass, low-redshift clusters detected by \textit{Planck} is not an accurate description of reality, but that is has a much larger amplitude than we measure from the real \textit{Planck} data.

Finally, we note that for the real re-extracted fixed $\hat{y}_0$ measurements there is excellent agreement between iMMF6 and iMMF5 (Figure\,\ref{fig:fixed_data}). This agreement implies that the $\sim 10$\% absolute calibration uncertainty of the \textit{Planck} 857\,GHz channel \citep{Planck2016VIII} has a negligible impact on our catalogues. This is something that we have verified for the first time regarding cluster detection, it not having been assessed in the official \textit{Planck} analyses, which only used all HFI channels (e.g., \citealt{Planck2012I,Planck2013XXIX,Planck2016xxvii}).


\section{Impact of the tSZ relativistic corrections}\label{sec:rsz}

The \textit{Planck} \texttt{SZiFi} catalogues have been produced ignoring the relativistic corrections to the tSZ SED, i.e., assuming that the electrons in the intracluster medium (ICM) of the detected clusters move at non-relativistic speeds. Indeed, in our MMFs, the tSZ SED is taken to be the non-relativistic SED, which is independent of electron temperature (see Section\,\ref{subsec:mmf}). However, it is well known that electrons in the ICM of massive clusters, such as those detectable by \textit{Planck}, have typical temperatures of a few keV \citep[e.g.,][]{Refregier2000, Nagai2007}. This implies that they move at a fraction of the speed of light, making the tSZ relativistic corrections potentially relevant to cosmological analyses \citep[e.g.,][]{Remazeilles2019, Rotti2021Giants}.

In this section we investigate the impact of this assumption on our catalogues. Specifically, we quantify the impact of the tSZ relativistic corrections on the cluster observables $\hat{y}_0$ and $q_{\mathrm{obs}}$. In addition, we assess whether the tSZ relativistic corrections contribute in any significant way to the small shifts observed at large $\hat{y}_0$ values in our re-extracted fixed $\hat{y}_0$ measurements for the CIB subsample (see Section\,\ref{sec:cib}). Our methods are presented in Section\,\ref{subsec:temperatures}, followed by our results in Section\,\ref{subsec:rsz_results}.

\subsection{Obtaining cluster temperatures and re-extracting $\hat{y}_0$}\label{subsec:temperatures}

We re-extract $\hat{y}_0$ and $q_{\mathrm{obs}}$ for the clusters in the CIB subsample with a redshift measurement (referred to as the `rSZ subsample') assuming the relativistic tSZ SED. There are 244 clusters in this rSZ subsample out of a total of 247 clusters in the CIB subsample. We use the same re-extraction pipeline as in our assessment of CIB contamination (Section\,\ref{sec:cib}), except that the tSZ SED is no longer non-relativistic.

Since the relativistic tSZ SED depends on the electron temperature \citep{Itoh1998,Challinor1998,Sazonov1998}, we first need a temperature estimate for each cluster. We obtain these temperature estimates using our blind iMMF6 $\hat{y}_0$ and $\hat{\theta}_{500}$ measurements and a set of scaling relations. Specifically, we use the following Compton-$y$--temperature scaling relation from \citet{Lee2022}, which has been calibrated with hydrodynamical simulations\footnote{We thank Elizabeth Lee for providing the fit parameter values for averaging over $R_{500}$, which were not included in \citet{Lee2022}. We have also corrected a typo in the normalisation of Eq.\,(16) of \citet{Lee2022}.}:
\begin{equation}\label{eq:t_scalrel}
T_Y = A(z) \left( \frac{Y_{500,\mathrm{physical}}}{10^{-5} \,\mathrm{Mpc}^2} \right)^{B(z)} E(z)^{2/5} \,\mathrm{keV}.
\end{equation}
Here, $T_Y$ is the Compton-$y$-weighted electron temperature, which is the most suitable integrated temperature regarding the tSZ relativistic corrections \citep{Lee2020,Lee2022}; $Y_{500,\mathrm{physical}}$ is the cluster's Compton-$y$ parameter integrated within the cluster's angular scale $\theta_{500}$, in physical units; and $E(z) = H(z)/H_0$ is the dimensionless expansion rate. The amplitude and exponent parameters, $A(z)$ and $B(z)$, are evaluated at each cluster's redshift by linearly interpolating the values obtained in \citet{Lee2022} at four redshifts, $\bmath{z} = (0, 0.25, 0.5, 1, 1.5)$, at which $A(z)$ and $B(z)$ were found to be $\bmath{A} = (3.123, 2.839, 2.584, 2.113)$ and $\bmath{B} = (0.364, 0.369, 0.363, 0.361)$, respectively. 

For the \citet{Arnaud2010} profile assumed for cluster detection and for given values of $y_0$ and $\theta_{500}$, the cluster's Compton-$y$ parameter integrated within $\theta_{500}$ in angular units, $Y_{500,\mathrm{angular}}$, can be written as 
\begin{equation}\label{eq:inty_angular}
Y_{500,\mathrm{angular}} = y_0  I \pi \theta_{500}^2, 
\end{equation}
where $I=0.06728$ is a numerical factor that is independent of the cosmological model, the cluster mass and redshift, and that we compute numerically. In physical units, the integrated parameter, $Y_{500,\mathrm{physical}}$, is then given by 
\begin{equation}\label{eq:inty_physical}
Y_{500,\mathrm{physical}} = Y_{500,\mathrm{angular}} D_{\mathrm{A}}^2 \left[\pi / (60 \times 180) \right]^2,
\end{equation}
where $D_{\mathrm{A}}$ is the angular diameter distance to the cluster's redshift, and where we note that this expression is valid if $Y_{500,\mathrm{angular}}$ is expressed in arcmin$^2$.

\begin{figure}
\centering
\includegraphics[width=0.4\textwidth,trim={0mm 0mm 0mm 0mm},clip]{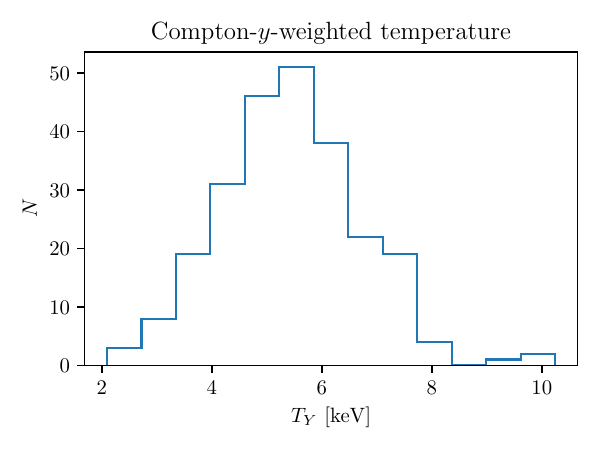}
\caption{Distribution of the estimated Compton-$y$-weighted electron temperatures $T_Y$ for our rSZ cluster subsample, which contains a total of 244 clusters (see Section\,\ref{sec:rsz}).}
\label{fig:temperature}
\end{figure}

For each cluster in the rSZ sample, we obtain an estimate for $Y_{500,\mathrm{physical}}$ using Eqs.\,(\ref{eq:inty_angular}) and\,(\ref{eq:inty_physical}). We use the cluster's blind $y_0$ and $\theta_{500}$ measurements, as obtained with iMMF6, since this MMF provides the highest signal-to-noise. As we show with synthetic observations in Section\,\ref{sec:validation}, these $Y_{500,\mathrm{physical}}$ estimates are highly correlated with their true values, although they present a Malmquist-type bias that we ignore here. Any differences between the true cluster profile and the \citet{Arnaud2010} profile that we assume will cause an additional bias. We then use the scaling relation of Eq.\,(\ref{eq:t_scalrel}) in order to obtain an estimate of the Compton-$y$-weighted temperature for each cluster. We evaluate $E(z)$ and $D_{\mathrm{A}}$ assuming the same cosmological model that was used to compute the cluster comoving distances in Section\,\ref{subsec:sky}.

The temperature distribution that we obtain for the rSZ subsample is shown in Figure\,\ref{fig:temperature}. As expected for the low-redshift, massive clusters that \textit{Planck} detects, the ICM electrons have temperatures of a few keV, with a sample mean temperature of $\left\langle T_{Y} \right\rangle= 5.437 \pm 0.082\,{\rm keV}$ that is consistent with the one found in \citet{Erler2018}, where we note that the quoted uncertainty is the empirical standard deviation on the mean. We note, however, that these temperature estimates must not be interpreted as accurate measurements, rather as estimates, better than the zero-temperature assumption underlying the non-relativistic tSZ SED. Indeed, they are obtained assuming a specific cluster profile and a simulation-calibrated Compton-$y$--temperature scaling relation, and are expected to be Malmquist-biased. Therefore, we have not attempted to marginalise over any measurement or modelling uncertainties and have not assigned measurement errors to them.

We then re-extract $\hat{y}_{0}$ and $q_{\mathrm{obs}}$ with the re-extraction pipeline of Section\,\ref{sec:cib}, i.e., using the MMF-averaged sky position and a fixed angular scale of $\theta_{500} = 5$\,arcmin. However, for each cluster we now construct the MMFs with the relativistic tSZ SED evaluated at the cluster's $T_Y$ estimate. We compute the relativistic tSZ SED with the \texttt{SZpack}\footnote{\texttt{\hyperref[https://github.com/CMBSPEC/SZpack]{github.com/CMBSPEC/SZpack}}} package \citep{Chluba2012SZpack,Chluba2013}, which we have interfaced with \texttt{SZiFi}. In particular, we use a combination of analytic approximations at the low temperature ($\lesssim 3\,{\rm keV}$) end, while above this temperature a pre-computed basis is used in order to avoid convergence issues inherent to the asymptotic expansion approach (see \citealt{Chluba2012SZpack} for a detailed discussion).

\subsection{Results and discussion}\label{subsec:rsz_results}

\begin{figure}
\centering
\includegraphics[width=0.5\textwidth,trim={0mm 0mm 0mm 0mm},clip]{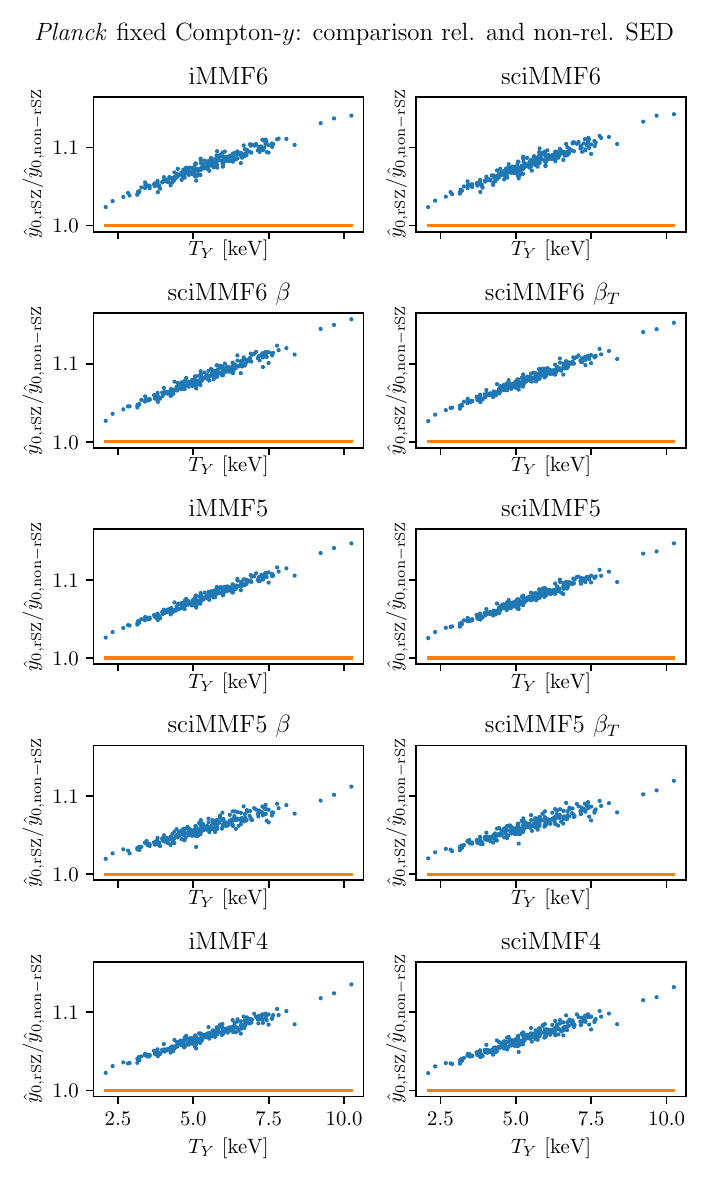}
\caption{Ratio between the measurements of the MMF amplitude parameter $\hat{y}_0$ for the rSZ sample extracted assuming a relativistic tSZ SED and their non-relativistic counterparts, as a function of the Compton-$y$-weighted electron temperature estimate $T_Y$, for our 10 MMFs. As expected, assuming a relativistic tSZ SED leads to $\hat{y}_0$ measurements that are greater than their non-relativistic counterparts, an effect that increases with the assumed temperature $T_Y$ (see Section\,\ref{subsec:rsz_results}).}
\label{fig:fixed_rsz_yt}
\end{figure}

\begin{figure}
\centering
\includegraphics[width=0.5\textwidth,trim={0mm 0mm 0mm 0mm},clip]{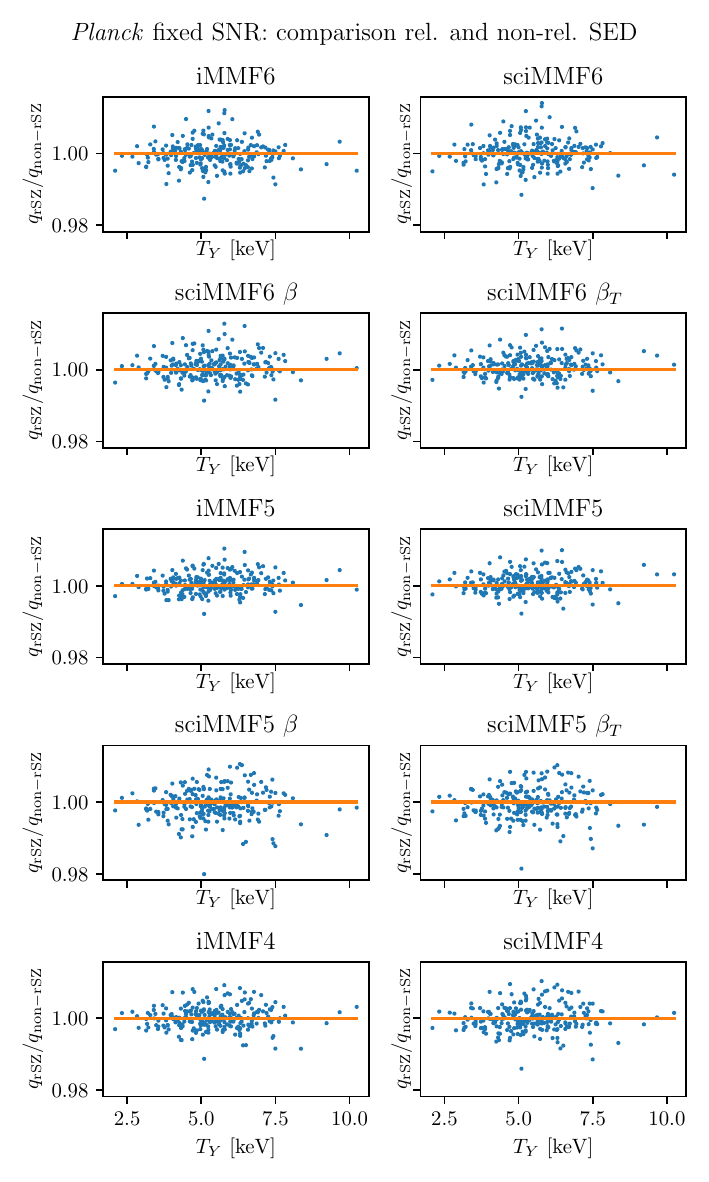}
\caption{As Figure\,\ref{fig:fixed_rsz_yt}, but for the signal-to-noise $q_{\mathrm{obs}}$. Assuming a relativistic tSZ SED produces $q_{\mathrm{obs}}$ values that are fully in agreement with those obtained with the non-relativistic SED. The impact of the tSZ relativistic corrections on the signal-to-noise measurements in our catalogues is therefore negligible. }
\label{fig:fixed_rsz_qt}
\end{figure}

Figure\,\ref{fig:fixed_rsz_yt} shows the ratio between the re-extracted fixed $\hat{y}_0$ measurements obtained with the relativistic tSZ SED and their non-relativistic counterparts, plotted against the electron temperature $T_Y$ estimates for the rSZ subsample and all our MMFs. As expected, this ratio increases with $T_Y$. Also as expected, the relativistic $\hat{y}_0$ measurements take larger values than their non-relativistic counterparts. This is due to the fact that, at the \textit{Planck} HFI frequencies, both below and above the tSZ null, the relativistic rSZ SED has a lower amplitude than the non-relativistic one, as comparatively more photons are scattered into higher frequencies, above 857\,GHz. At a fixed signal, this decrease in the amplitude of the SED is compensated by a larger estimated Compton-$y$ value (see, e.g., \citealt{Remazeilles2019}).

The shifts in the $\hat{y}_0$ measurements seen in Figure\,\ref{fig:fixed_rsz_yt} are at the $5$--$10$\,\% level for most clusters across all MMFs, indicating that the $\hat{y}_0$ measurements in our catalogues are probably biased by about $5$--$10$\,\% due to the assumption of the non-relativistic tSZ SED for cluster detection (a bias that adds to the other biases mentioned above). Interestingly, these shifts only depend mildly on the MMF used, at the $1$--$2$\,\% level.

Despite $\hat{y}_0$ changing by about $5$--$10$\,\%, the signal-to-noise $q_\mathrm{obs}$ exhibits no systematic shift at all.\footnote{A similar conclusion has also been reached by A. Rotti (private communication).}
This can be seen in Figure\,\ref{fig:fixed_rsz_qt}, which is the signal-to-noise analogue to Figure\,\ref{fig:fixed_rsz_yt}. In it it is apparent that the $q_\mathrm{obs}$ values are insensitive to the change in the tSZ SED for all MMFs, with any shifts being constrained to less than 1\,\% in $q_\mathrm{obs}$. We recall that our temperature estimates are not to be regarded as accurate measurements, but rather as estimates. Their range, however, spans over a factor of 4 in temperature, and no temperature dependency is seen in Figure\,\ref{fig:fixed_rsz_qt}. It follows that our findings regarding the signal-to-noise would not change significantly if our temperature estimates were, say, a factor of 2 away from the true electron temperatures. We can, therefore, safely conclude that the tSZ relativistic corrections have a negligible impact on the signal-to-noise measurements in our catalogues, even if the inferred $\hat{y}_0$ values are noticeably biased. This result is particularly relevant for cosmological number-count analyses, in which the signal-to-noise is typically used as the cluster tSZ observable. We note that the tSZ relativistic corrections will, however, add some scatter to the signal-to-noise values, which could be decreased through the use of a relativistic SED for cluster extraction, as performed here for the tSZ subample.

\begin{figure}
\centering
\includegraphics[width=0.5\textwidth,trim={0mm 0mm 0mm 0mm},clip]{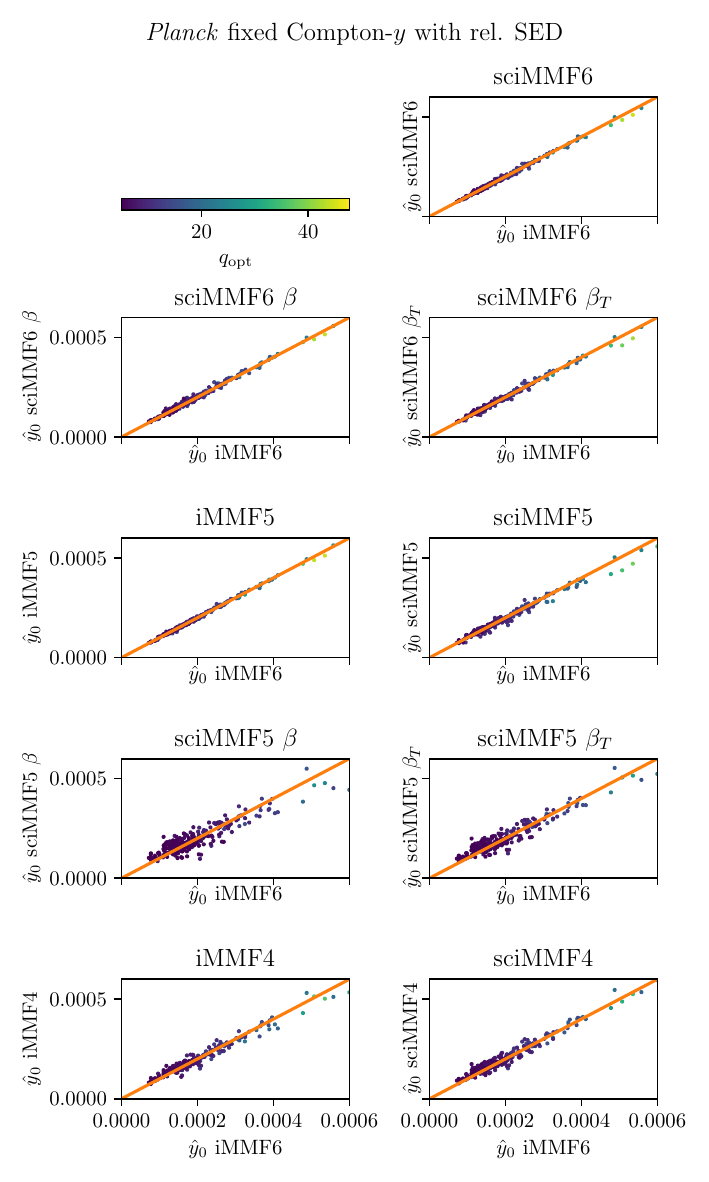}
\caption{As Figure\,\ref{fig:fixed_data}, but for the re-extracted fixed $\hat{y}_0$ measurements obtained assuming a relativistic tSZ SED. Note that there are barely any changes with respect to Figure\,\ref{fig:fixed_data}, a consequence of $\hat{y}_0$ changing by a very similar amount for all MMFs (Figure\,\ref{fig:fixed_rsz_yt}).}
\label{fig:fixed_rsz}
\end{figure}

Finally, Figure\,\ref{fig:fixed_rsz} is analogous to Figure\,\ref{fig:fixed_data}, comparing the re-extracted fixed $\hat{y}_0$ measurements for each MMF with those for iMMF6, but here for the measurements obtained assuming the relativistic tSZ SED. It is virtually identical to Figure\,\ref{fig:fixed_data}. This was expected given the minimal variation in the shifts of the $\hat{y}_0$ measurements across different MMFs seen in Figure\,\ref{fig:fixed_rsz_yt}. Thus, the tSZ relativistic corrections seem an unlikely explanation for the small shifts in the $\hat{y}_0$ measurements at large $\hat{y}_0$ values observed for some MMFs.

This is the first time that the significance of the tSZ relativistic corrections on the cluster signal-to-noise measurements of tSZ-selected samples is quantified with real data. Indeed, this issue was not investigated in the official \textit{Planck} analyses (e.g., \citealt{Planck2012I,Planck2013XXIX,Planck2016xxvii}), nor for the ACT and SPT catalogues \citep{Bleem2015,Hilton2018,Bleem2020,Bleem2022,Bleem2024}. In \citet{Hilton2018}, a scaling relation including a term accounting for the tSZ relativistic corrections was proposed, but its significance was not quantified with data.

\section{Validation of the cluster detection pipeline}\label{sec:validation}

In this section we describe how we validate our \texttt{SZiFi}-based cluster detection pipeline with synthetic data. In particular, we apply it to synthetic \textit{Planck}-like frequency maps, producing a `validation catalogue' for each of the 10 MMFs considered in our analysis of the \textit{Planck} data (Section\,\ref{subsec:synthetic_data}; see also Section\,\ref{subsec:matched_filters_used}). We place a particular emphasis in testing our ability to theoretically predict the cluster counts in our validation catalogues as a function of signal-to-noise and redshift (Section\,\ref{subsec:number_counts_sim}), as this is an important step forward towards using our real catalogues in a cosmological number-count analysis. We also quantify the purity of the validation catalogues, assessing the number and nature of false detections (Section\,\ref{subec:purity_sim}), and investigate how the inferred cluster observables ($\hat{y}_0$, $\hat{\theta}_{500}$, integrated Compton-$y$, and cluster sky coordinates) are related to their true values (Sections\,\ref{subsec:obs_sim} and \ref{subsec:sky_location_sim}). Throughout this section, we focus on our validation iMMF6 and MMF6 catalogues, analysing the other validation catalogues in Appendix\,\ref{appendix:validation}.

\subsection{Synthetic maps and cluster catalogues}\label{subsec:synthetic_data}

\subsubsection{Synthetic input catalogue}

We generate a synthetic cluster catalogue by obtaining samples from the halo mass function using the synthetic catalogue generator of the \texttt{cosmocnc}\footnote{\texttt{\href{https://github.com/inigozubeldia/cosmocnc/tree/main}{github.com/inigozubeldia/cosmocnc}}} package \citep{Zubeldia2024,Zubeldia2024Grenoble}. We use the \citet{Tinker2008} halo mass function, computed within \texttt{cosmocnc} using the \texttt{cosmopower}\footnote{ \href{https://github.com/cosmopower-organization}{\texttt{github.com/cosmopower-organization}}}  power spectra emulator \citep{Spurio2022,Bolliet2023}, and assume a flat $\Lambda$CDM cosmological model with $\Omega_{\mathrm{c}} = 0.26603$, $\Omega_{\mathrm{b}} = 0.04897$, $h = 0.674$, $A_s= 2.08467 \times 10^{-9}$, $n_s = 0.96$, and $\sum m_\nu = 0.06$\,eV. The parameter values are taken to be consistent with the \textit{Planck} CMB constraints \citep{Planck2018VI}. We assume a minimum mass of $M_{500,\mathrm{min}} = 4 \times 10^{14} M_{\odot}$ and a maximum mass of $M_{500,\mathrm{max}} = 10^{16} M_{\odot}$, and minimum and maximum redshifts $z_{\mathrm{min}} = 0.01$ and $z_{\mathrm{max}} = 1$, respectively. The redshift limits and the lower mass limit are chosen so that most clusters to be detected in \textit{Planck} data are spanned (see, e.g., Figure\,1 of \citealt{Ade2016}).

The synthetic catalogue is obtained by first calculating the mean total number of clusters in the full sky within the mass and redshift range considered, $\bar{N}_{\mathrm{tot}}$. This is done by numerically integrating the halo mass function multiplied by the differential volume element. The actual realisation of the total number of clusters, $N_{\mathrm{tot}}$, is then obtained as a Poisson draw from $\bar{N}_{\mathrm{tot}}$. We obtain $N_{\mathrm{tot}} = 8683$. A total of 8683 mass--redshift pairs are then drawn from the product of the halo mass function with the differential volume element, and a random sky position is then assigned to each of them. We refer the reader to \citet{Zubeldia2024} for further details about the catalogue generator.

\subsubsection{Synthetic sky maps}

We use our synthetic cluster catalogue to produce a full-sky Compton-$y$ map by painting in a Compton-$y$ profile for each cluster. We assume the same cluster profile that is used in our MMFs for cluster finding, namely the Compton-$y$ profile due to the \citet{Arnaud2010} pressure profile truncated at $5\theta_{500}$. We evaluate the profile at its best-fit parameter values and at a biased mass $\beta_{\mathrm{SZ}} M_{500}$, where $\beta_{\mathrm{SZ}} = 0.62$ is the so-called `hydrostatic mass bias' (hereafter, the `SZ mass bias'), whose value we take to be consistent with that which is found to reconcile the \textit{Planck} MMF3 cosmology sample with the \textit{Planck} CMB  \citep{Ade2016}. We use a HEALPix pixellation with $N_{\mathrm{SIDE}} = 2048$ that is identical to that of the \textit{Planck} HFI maps.

Next, we generate a synthetic tSZ signal map for each observation frequency channel. We do this by convolving our Compton-$y$ map with the instrument beam for each channel and rescaling it by the non-relativistic tSZ SED at each channel effective frequency. We assume six frequency channels at 100, 143, 217, 353, 545, and 857\,GHz, taking the band transmission profiles to be delta functions and Gaussian beams with FWHMs of 9.68, 7.3, 5.02, 4.94, 4.83, and 4.64\,arcmin (taken from \citealt{Planck2016VIII}).

We then add noise and beam-convolved foregrounds to our synthetic tSZ maps. We use the same synthetic \textit{Planck}-like noise and foreground maps that were used in Section\,\ref{sec:cib}, recalling that the latter comprise three foregrounds: the lensed CMB, the kSZ and the CIB, all taken from the Websky simulation. By construction, the CIB maps are not spatially correlated with our synthetic tSZ signal.

\subsubsection{Validation cluster catalogues}

We apply our \texttt{SZiFi}-based cluster finding pipeline to our synthetic sky maps. We use exactly the same configuration that was used in the processing of the real \textit{Planck} maps (see Section\,\ref{sec:detection}) except in the modelling of the beams (using Gaussian beams instead of the RIMO ones) and of the band transmission profiles (assuming delta function profiles). We use the same Galactic mask that was employed for our analysis of the real \textit{Planck} maps (the \textit{Planck} PR2 80\,\% Galactic mask) and the point-source mask that we constructed for the Websky CIB in Section\,\ref{sec:cib}. We impose a final signal-to-noise threshold of $q_{\mathrm{th}} = 5$.

We produce a non-iterative and an iterative catalogue for each of the 10 MMFs considered in our real analysis (see Section\,\ref{subsec:matched_filters_used}), referring to these catalogues as our validation catalogues. We then cross-match each validation catalogue with the input synthetic catalogue. We use the cross-matching algorithm described in Section\,\ref{subsec:master_catalogue}, which here is used to assess whether each detection corresponds to a real cluster (if it has a counterpart in the input catalogue) or is a false detection (if it has no counterpart).

As for our real catalogues (see Section\,\ref{subsec:inspection}), after cross-matching the validation catalogues with the input catalogue we manually inspect all the unconfirmed detections, removing what we deem to be true false detections. As discussed in Section\,\ref{subsec:inspection}, these are (i) large-angular-scale detections located very close to the tile edges that have not been successfully merged with the corresponding detection in the neighbouring tile with which they form a single object, or (ii) large-angular-scale detections very close to point-source-masked regions. For reference, we find 5 detections of the first kind and 3 of the second kind out of a total of 42 unconfirmed detections in our validation iMMF6 catalogue, 5 of the first kind and 4 of the second kind out of a total of 44 in our validation iMMF5 catalogue, and 6 of the first kind and 3 of the second kind out of a total of 62 in our validation iMMF4 catalogue. These numbers are comparable to those in our real \textit{Planck} catalogues (see Section\,\ref{subsec:inspection}).

\subsection{Number counts}\label{subsec:number_counts_sim}

We first consider the number counts in our validation catalogues as a function of signal-to-noise and redshift, assessing whether we can produce an accurate theoretical prediction for them. We compute our theoretical number-count prediction with the \texttt{cosmocnc} package, which for a given cluster mass observable (e.g., the tSZ signal-to-noise) can evaluate the cluster number counts as a function of the observable and of redshift. We assume the same halo mass function that we used to generate our input catalogue, and evaluate it at the input cosmology.

Our mass observable is the tSZ signal-to-noise, $q_{\mathrm{obs}}$. Following \citet{Zubeldia2021}, its expected value at given mass $M_{500}$ and redshift $z$, $\bar{q}_{\mathrm{obs}} (M_{500},z)$, can be written as
\begin{equation}\label{eq:snr_model}
    \bar{q}_{\mathrm{obs}} (M_{500},z) = \left[ \bar{q}^2 (M_{500},z) + f \right]^{1/2}\,,
\end{equation}
where $\bar{q} (M_{500},z)$ is the mean signal-to-noise extracted at the true sky position and angular size of the cluster and $f$ is the number of degrees of freedom over which the signal-to-noise is maximised in the cluster detection process. There are $f=3$ degrees of freedom: the cluster's two sky coordinates and its angular scale $\theta_{500}$. As discussed in \citet{Zubeldia2021}, neglecting this degree-of-freedom correction by setting $f = 0$ (as done, e.g., in \citealt{Ade2016}) leads to an `optimisation bias' in the modelling of the clusters' signal-to-noise measurements. For the \citet{Arnaud2010} pressure profile $\bar{q} (M_{500},z)$ can be related to the cluster mass and redshift with the scaling relation
\begin{equation}\label{eq:snr_model2}
    \bar{q} (M_{500},z) = \frac{y_0 ( \beta_{\mathrm{SZ}} M_{500},z)}{\sigma_{\mathrm{f}} (\theta_{500}  (\beta_{\mathrm{SZ}} M_{500},z))}\,,
\end{equation}
where $y_0$ is the central Compton-$y$ value of the cluster and $\sigma_{\mathrm{f}}$ is the MMF noise evaluated at the cluster's angular scale $\theta_{500}$ (see Section\,\ref{subsec:mmf}). $y_0$, in turn, is given by
\begin{equation}\label{eq:scalrelsz}
    y_0 ( \beta_{\mathrm{SZ}} M_{500},z) = 10^{A_{\mathrm{SZ}}} \left( \frac{  \beta_{\mathrm{SZ}} M_{500}}{ 3 \times 10^{14} h_{70}^{-1} M_{\odot}} \right)^{\alpha_{\mathrm{SZ}}} E^2(z) h_{70}^{-1/2}\,,
\end{equation}
where $A_{\mathrm{SZ}}=-4.3054$, $\alpha_{\mathrm{SZ}}=1.12$, $h_{70} = h/0.7$ and $E(z) = H(z)/H_0$. All the cosmology-dependent quantities used to compute $\theta_{500}$ and $E(z)$ are evaluated at the input cosmology. 

Since there is no intrinsic scatter, as the only source of scatter in our synthetic maps is that due to the instrumental noise and foregrounds, within the \texttt{cosmocnc} framework $q_{\mathrm{obs}}$ is said to be linked to the cluster mass and redshift with a one-layer hierarchical model. In the only layer, the mass observable $q_{\mathrm{obs}}$ has a mean given by Eq.\,(\ref{eq:snr_model}) and unit-variance Gaussian scatter, as the signal-to-noise is defined by normalising the MMF amplitude parameter $\hat{y}_0$ by its variance (see Eq.\,\ref{eq:snr_def}).

\begin{figure}
\centering
\includegraphics[width=0.5\textwidth,trim={0mm 0mm 0mm 0mm},clip]{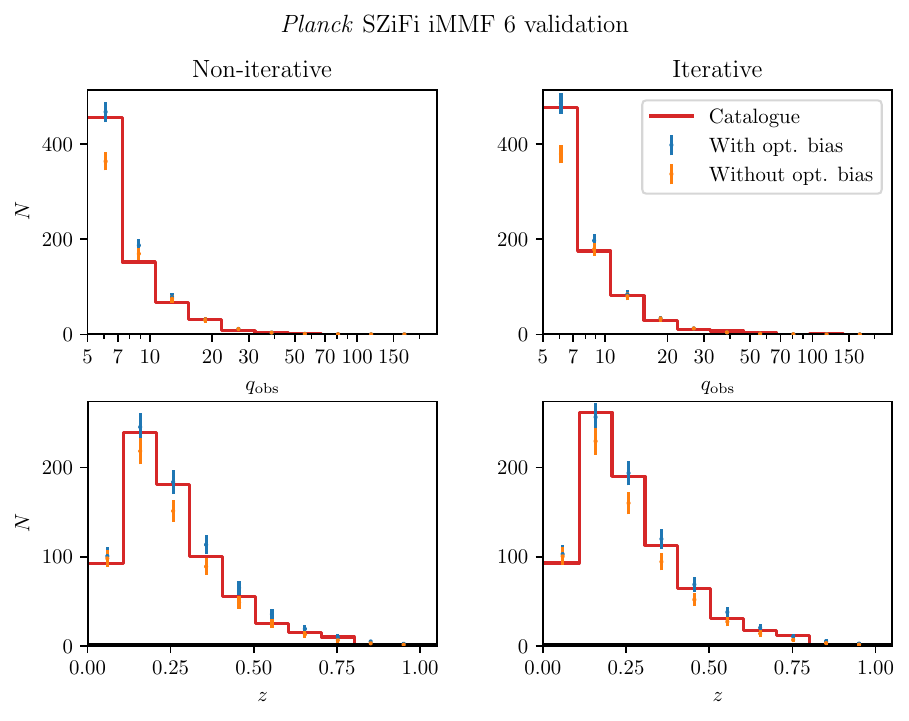}
\caption{In red, number counts in our validation MMF6 and iMMF6 catalogues (left and right panels, respectively) as a function of both signal-to-noise and redshift (upper and lower panels, respectively); only confirmed (i.e., true) detections are shown. Theoretical predictions for the counts are shown as the blue data points (if the optimisation bias is properly taken into account) and as the orange data points (if it is neglected) with the error bars corresponding to their Poisson errors. If the optimisation bias is taken into account the theoretical prediction provides an excellent description of the observed counts for the iterative (iMMF6) catalogue across both signal-to-noise and redshift.}
\label{fig:validation_counts}
\end{figure}

Figure\,\ref{fig:validation_counts} presents histograms of the number counts as a function of $q_{\mathrm{obs}}$ (upper panels) and redshift (lower panels) for our MMF6 (non-iterative) and iMMF6 (iterative) catalogues (left and right panels, respectively). Only the true detections, i.e., those with a counterpart in the input catalogue, are shown. Figure\,\ref{fig:validation_counts} also shows the corresponding \texttt{cosmocnc} predictions made assuming the input halo mass function and the scaling relation of Eqs.\,(\ref{eq:snr_model}--\ref{eq:scalrelsz}), both if the optimisation bias is accounted for by setting $f=3$ (blue points) and also if it is ignored by setting $f=0$ instead (orange points). For each predicted value we also show the associated Poisson error bar, given by its square root. We stress that these are predictions and not fits to the observed counts.

As it is apparent in Figure\,\ref{fig:validation_counts}, our prediction with the optimisation bias taken into account successfully describes the number counts in the iMMF6 validation catalogue, both across signal-to-noise and redshift. For its non-iterative counterpart, the observed counts fall systematically below the predicted ones by a small amount ($\sim 1\,\sigma$). This is due to the negative covariance bias in the signal-to-noise measurements that we briefly discuss in Section\,\ref{subsec:iterative} and is investigated in detail in \citet{Zubeldia2022}. Note, in particular, that this effect is not due to the noise covariance in the non-iterative case being overestimated, as the same (overestimated) noise covariance is used in the theoretical prediction (in particular, in Eq.\,\ref{eq:snr_model2}).

It is also clear in Figure\,\ref{fig:validation_counts} that the optimisation bias is a very significant effect a the low signal-to-noise end. While adopting the simple degree-of-freedom correction of Eq.\,(\ref{eq:snr_def}) results in a good prediction for the observed counts, neglecting the optimisation bias leads to a significant under prediction of the observed counts. This number-count under prediction at low signal-to-noise spreads over a broad redshift range, as can be seen in the lower panels.

Similar conclusions can be drawn by simply considering the total number of clusters in the catalogue. The iterative catalogue contains a total of 785 clusters. If the optimisation bias is taken into account $818 \pm 29$ clusters are predicted (1.14\,$\sigma$ away from the observed value), whereas if it is ignored, $688 \pm 26$ clusters are predicted instead ($-3.73$\,$\sigma$ away from the observed value). On the other hand, the non-iterative catalogue consists of 719 clusters. Its corresponding theoretical predictions are $780 \pm 28$ if the optimisation bias is taken into account, and $654 \pm 25$ otherwise, values which are, respectively, 2.17\,$\sigma$ and $-2.60$\,$\sigma$ away from the observed value.

In summary, when the noise covariance is estimated iteratively and the optimisation bias is taken into account, our model successfully describes the observed counts in our validation catalogue (iMMF6 catalogue) as a function of both signal-to-noise and redshift. This constitutes an important stepping stone towards using our real iMMF6 catalogue for cosmological inference. In Appendix\,\ref{appendix:validation} we consider our remaining validation catalogues, similarly finding good agreement between the observed counts and their theoretical prediction (see, in particular, Figure\,\ref{fig:validation_counts_all}).

\subsection{Purity}\label{subec:purity_sim}

\begin{figure}
\centering
\includegraphics[width=0.4\textwidth,trim={0mm 0mm 0mm 0mm},clip]{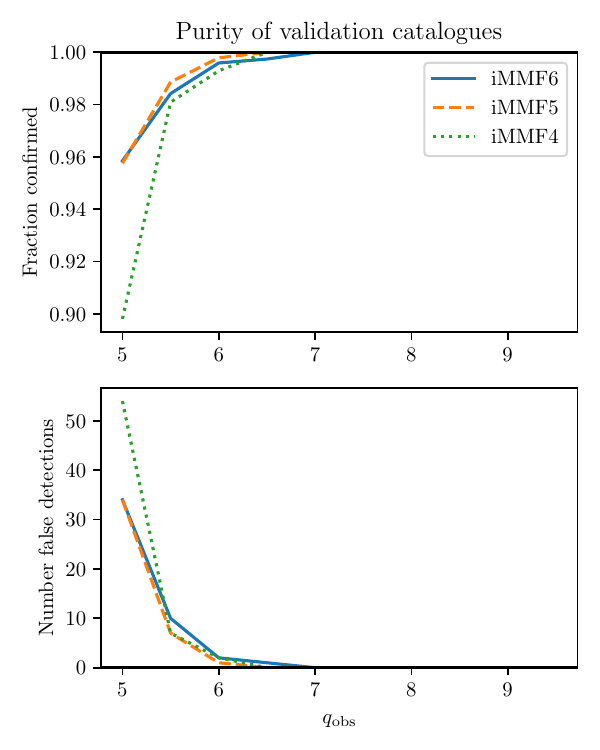}
\caption{\textit{Upper panel}: Fraction of true detections (or purity) in our validation catalogues as a function of the signal-to-noise threshold $q_{\mathrm{th}}$. \textit{Lower panel}: Associated number of false detections. As expected, the purity increases quickly with $q_{\mathrm{th}}$, being 100\,\% at $q_{\mathrm{th}} > 7$.}
\label{fig:validation_purity}
\end{figure}

The upper panel of Figure\,\ref{fig:validation_purity} shows the fraction of confirmed detections in our validation iMMF6, iMMF5, and iMMF4 catalogues as a function of signal-to-noise threshold. As the catalogue against which the validation catalogues have been cross-matched is the input cluster catalogue, this fraction corresponds to the fraction of true detections in the catalogue, also known as catalogue purity. The lower panel, on the other hand, shows the associated number of unconfirmed detections, which in this case corresponds to the number of false detections.

The overall picture is similar to that for our real \textit{Planck} catalogues (see Section\,\ref{subsec:validation}). The purity increases quickly with signal-to-noise from about 96\,\% for iMMF6 and iMMF5 and about 90\,\% for iMMF4 at $q_{\mathrm{th}}=5$ to over 99\,\% for all catalogues at $q_{\mathrm{th}}=6$ and 100\,\% for all catalogues at $q_{\mathrm{th}}=7$. At $q_{\mathrm{th}}=5$, the purity for iMMF4 is significantly lower than for iMMF5 and iMMF6, but this difference is reduced significantly at $q_{\mathrm{th}}=5.5$.

\subsection{Cluster observables}\label{subsec:obs_sim}

\begin{figure}
\centering
\includegraphics[width=0.5\textwidth,trim={0mm 0mm 0mm 0mm},clip]{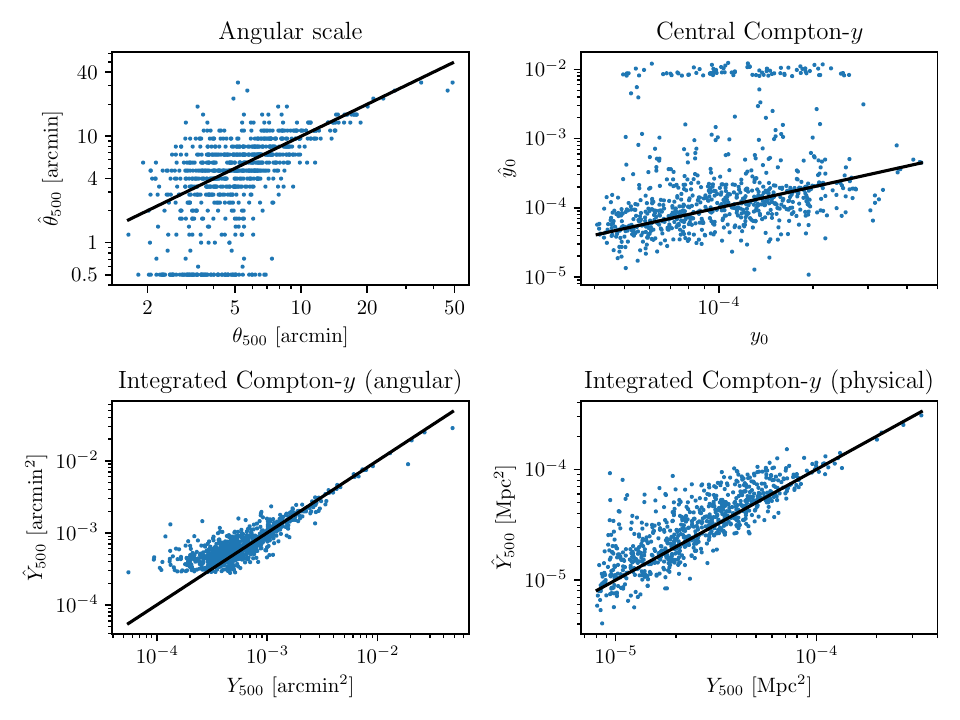}
\caption{Inferred cluster observable values plotted against their corresponding true values in the iMMF6 validation catalogue. The observables shown are the cluster angular scale $\theta_{500}$ (upper left panel), the cluster central Compton-$y$ $y_0$ (which acts as the MMF amplitude parameter; upper right panel), and the cluster integrated Compton-$y$ parameter, both in angular and physical units ($Y_{500,\mathrm{angular}}$ and $Y_{500,\mathrm{physical}}$; lower left and right panels, respectively).}
\label{fig:validation_input_output}
\end{figure}

Figure\,\ref{fig:validation_input_output} shows the inferred values of four cluster observables for the true detections in our validation iMMF6 catalogue (vertical axes) plotted against their true values in the synthetic data (horizontal axes). In particular, the upper left panel corresponds to the cluster angular scale, whose inferred value for each cluster, $\hat{\theta}_{500}$, corresponds to the MMF angular scale at which the signal-to-noise is maximised. The inferred angular scale is found to be correlated with its true value, but with significant scatter. This is a consequence of the signal-to-noise depending weakly on the angular scale for low-significance detections that dominate the sample. Notably, there is a set of clusters that are detected at the smallest search angular scale, $\hat{\theta}_{500}=0.5$\,arcmin, and that appear as outliers in the $\hat{\theta}_{500}$--$\theta_{500}$ distribution. These detections correspond to clusters that are not resolved by the cluster-finding algorithm for which the signal-to-noise slowly keeps increasing for arbitrarily small values of $\hat{\theta}_{500}$.

The upper right panel is an analogous plot for the MMF amplitude, i.e., the central cluster Compton-$y$ value. Its inferred value, $\hat{y}_0$, which corresponds to the MMF amplitude evaluated at the inferred angular scale $\hat{\theta}_{500}$, is similarly found to correlate with its true value $y_0$. Here, there is also significant scatter, mostly caused by the scatter in $\hat{\theta}_{500}$. Indeed, all detections have a signal-to-noise greater than 5, which implies that the noise on  $\hat{y}_0$ at a given angular scale is $20$\,\% of its value or less. The outlier clusters of the upper left panel translate here into clusters with very large values of  $\hat{y}_0$, with $\hat{y}_0 \sim 10^{-2}$. This can be thought of as an artefact of the detection method: for unresolved clusters, as $\hat{\theta}_{500}$ goes to arbitrarily small scales (bound here by the lowest search angular scale), $\hat{y}_0$ takes arbitrarily large values, so that the signal-to-noise is (approximately) preserved, with it varying very slowly as a function of $\hat{\theta}_{500}$.

The lower panels are analogous plots for Compton-$y$ parameter integrated within $\hat{\theta}_{500}$, expressed both in angular units, $Y_{500,\mathrm{angular}}$ (lower left panel), and in physical units, $Y_{500,\mathrm{physical}}$ (lower right panel). For each cluster, we use Eqs.\,(\ref{eq:inty_angular}) and (\ref{eq:inty_physical}), which assume the \citet{Arnaud2010} profile, to obtain inferred values for $Y_{500,\mathrm{angular}}$ and $Y_{500,\mathrm{physical}}$, $\hat{Y}_{500,\mathrm{angular}}$ and $\hat{Y}_{500,\mathrm{physical}}$, from the corresponding $\hat{\theta}_{500}$ and $\hat{y}_0$ measurements.

As for $\hat{\theta}_{500}$ and $\hat{y}_0$, $\hat{Y}_{500,\mathrm{physical}}$, $\hat{Y}_{500,\mathrm{angular}}$ are found to correlate with their corresponding true values. For $\hat{Y}_{500,\mathrm{angular}}$, which can be thought of as being measured more directly by the MMF, a Malmquist-type bias can be clearly observed at the low $\hat{Y}_{500,\mathrm{angular}}$ end. This selection effect is also visible in the lower right panel, where it is spread over a broader $\hat{Y}_{500,\mathrm{physical}}$ range. Note that for both $\hat{Y}_{500,\mathrm{angular}}$ and $\hat{Y}_{500,\mathrm{physical}}$ there is significant scatter, which is dominated by the uncertainty in the angular scale, as for $\hat{y}_0$. Also note that for both integrated quantities there are no outliers, indicating that these quantities are much less affected than $\hat{\theta}_{500}$ and $\hat{y}_0$ by the artefact of the detection method described above.

Similar results are obtained for the other MMFs; we have omitted these for conciseness.

\subsection{Sky position uncertainty}\label{subsec:sky_location_sim}

\begin{figure}
\centering
\includegraphics[width=0.4\textwidth,trim={0mm 0mm 0mm 0mm},clip]{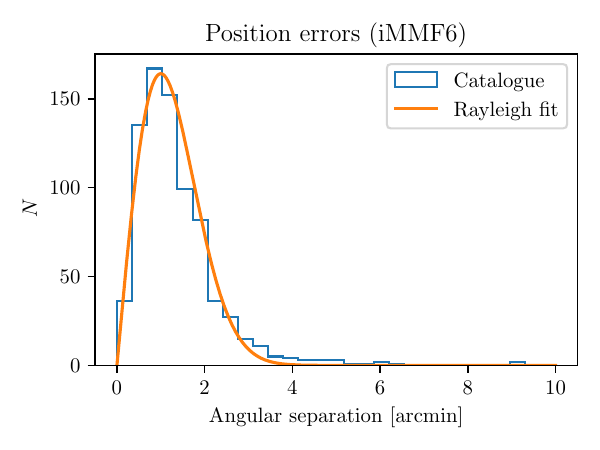}
\caption{Distribution of the angular distance between the inferred cluster sky position and the input value $\Delta\theta$ for the clusters in the iMMF6 validation catalogue (blue histogram) together with its Rayleigh distribution fit (orange curve), which has a scale parameter $\sigma_{\mathrm{misc}} = 1.001 \pm 0.018$.}
\label{fig:validation_position}
\end{figure}

We quantify the uncertainty in the inferred sky positions of the clusters in our validation catalogues by computing, for each cluster, the angular distance $\Delta \theta$ between the inferred position and its true sky position. We then model the resulting $\Delta \theta$ distribution (the `miscentering distribution') as a Rayleigh distribution for each catalogue, fitting for its scale parameter, denoted with $\sigma_{\mathrm{misc}}$. This model assumes that the uncertainty in the inferred values for the two cluster sky coordinates follows a Gaussian distribution. We perform the fit in a Bayesian way by assuming each cluster to be statistically independent from each other and to follow the same Rayleigh distribution. Under these assumptions, the likelihood for the $\Delta \theta$ vector $\bmath{\Delta} \bmath{\theta}$ at a given value of $\sigma_{\mathrm{misc}}$ can be written as
\begin{equation}\label{eq:misc}
P(\bmath{\Delta} \bmath{\theta} | \sigma_{\mathrm{misc}}) = \prod_i \mathcal{R} ( \Delta \theta_i ; \sigma_{\mathrm{misc}}),
\end{equation}
where $\mathcal{R} (\Delta \theta ; \sigma_{\mathrm{misc}})$ denotes a Rayleigh distribution with scale parameter $\sigma_{\mathrm{misc}}$ and the product runs over over all the clusters in the catalogue. Assuming a flat prior on $\sigma_{\mathrm{misc}}$, Eq.\,(\ref{eq:misc}) can be thought of a a posterior for $\sigma_{\mathrm{misc}}$. We use this posterior to infer the value of $\sigma_{\mathrm{misc}}$ by evaluating it on a $\sigma_{\mathrm{misc}}$ grid and, in turn, numerically integrating the evaluated posterior, obtaining the inferred mean and standard deviation of $\sigma_{\mathrm{misc}}$. We do this taking into account only the clusters with $\Delta \theta < 3.5$\,arcmin, as we find that the clusters with a larger positional uncertainty lead to significantly worse fits to the observed miscentering distribution.

Figure\,\ref{fig:validation_position} shows the observed miscentering distribution for the iMMF6 validation catalogue (blue histogram), together with the Rayleigh distribution fit for the mean of the inferred value of the scale parameter, which we find to be $\sigma_{\mathrm{misc}} = 1.001 \pm 0.018$ (orange curve). The Rayleigh distribution provides a reasonably good fit to the observed distribution, the only significant departure being a heavier tail in the observed distribution at large values of $\Delta \theta$ ($\Delta \theta \gtrsim 3$\,arcmin).

Similar fits are obtained for our other validation catalogues. The scale parameter values that we find are: $\sigma_{\mathrm{misc}} = 0.993 \pm 0.018$ (sciMMF6 $\beta$), $\sigma_{\mathrm{misc}} = 1.034 \pm 0.021§$ (sciMMF6 $\beta$), $\sigma_{\mathrm{misc}} = 1.048 \pm 0.020$ (sciMMF6 $\beta_T$), $\sigma_{\mathrm{misc}} = 1.013 \pm 0.019$ (iMMF5), $\sigma_{\mathrm{misc}} 1.023 \pm 0.020$ (sciMMF5), $\sigma_{\mathrm{misc}} = 1.003 \pm 0.036$ (sciMMF5 $\beta$), $\sigma_{\mathrm{misc}} = 1.040 \pm 0.035$ (sciMMF5 $\beta_T$), $\sigma_{\mathrm{misc}} = 1.029 \pm 0.024$ (iMMF4), $\sigma_{\mathrm{misc}} = 1.061 \pm 0.030$ (sciMMF4). Note that there is little variation in the miscentering scale parameter across our different filters.

We note that our procedure likely underestimates the position uncertainty, as the clusters in our synthetic data have the same spherically-symmetric profile. The positional errors should also be interpreted as being with respect to the true tSZ peak and not with respect to other cluster centres (e.g., the peak of the gravitational potential). We also note that more complex models for the miscentering distribution can be considered, e.g., models in which the distribution depends on the cluster signal-to-noise (see, e.g., \citealt{Hilton2018,Bocquet2023}). These models have the potential to provide a better fit to the observed distribution. We leave the exploration of these models to further work.

\section{Summary}\label{sec:summary}

In this work we have introduced the \textit{Planck} \texttt{SZiFi} galaxy cluster catalogues, a set of 10 catalogues obtained by applying the \texttt{SZiFi} tSZ cluster finder to the \textit{Planck} PR3 HFI temperature maps, detecting objects down to a signal-to-noise threshold of 5. Each catalogue is produced with a different frequency channel and CIB deprojection combination (see Table\,\ref{table:mmfs}). In particular, three of them, the iMMF6 (our baseline catalogue), iMMF5, and iMMF4 catalogues are constructed with a standard iterative MMF: iMMF6 using all HFI channels, iMMF5 excluding the highest-frequency channel, and iMMF4 excluding the two highest-frequency channels. The remaining seven catalogues are obtained for the same three channel combinations, but using \texttt{SZiFi}'s spectrally constrained MMFs instead, deprojecting the CIB and, for four of them, also the first-order moment with respect to the spectral parameter $\beta$ or $\beta_T$. In addition, each catalogue has a non-iterative counterpart that is obtained without iterative noise covariance estimation. We have compiled all the catalogues into a single \textit{Planck} \texttt{SZiFi} master catalogue containing a total of 1499 detections, which we have cross-matched with a set of six existing tSZ and X-ray-selected catalogues and meta-catalogues including a significant number of post-\textit{Planck} detections. The master catalogue will become publicly available at \href{https://github.com/inigozubeldia/szifi/tree/main/planck_szifi_master_catalogue}{this link} and a brief description of its entries can be found in Table\,\ref{table:catalogue}.

The \textit{Planck} \texttt{SZiFi} catalogues present two key improvements relative to the official \textit{Planck} catalogues. First, the MMFs with which they are obtained are constructed using an iterative estimate of the noise covariance. This is achieved by masking the detections made in a first run of the cluster finder from the input temperature maps and re-estimating the covariance with the masked maps. As was shown in \citet{Zubeldia2023} and is discussed here, using a non-iterative covariance estimate, as done in the official \textit{Planck} catalogues, leads to a bias in the cluster tSZ observables (signal-to-noise and Compton-$y$ estimate $\hat{y}_0$) and to a loss of signal-to-noise. Both effects disappear with our iterative procedure. The second key improvement is the deprojection of the CIB in 7 of the 10 \texttt{SZiFi} catalogues (both iterative and non-iterative) through the use of spectrally constrained MMFs. This constitutes the first application of spectrally constrained MMFs to real data and the first time that any foreground is deprojected in the context of tSZ cluster finding.

Our baseline catalogue, the iMMF6 catalogue, contains the largest number of detections, with 833 of them in total, 766 of which have an available redshift measurement. The redshift distribution peaks at about $z=0.2$ (see Figure\,\ref{fig:histograms}). It has an empirical lower bound on the purity of 95\% at a signal-to-noise of 5, 99\% at 6, and 100\% at 7 (see Figure\,\ref{fig:confirmation}). As fewer channels are used and/or more SEDs are deprojected, fewer objects are detected, a consequence of the expected associated signal-to-noise penalty (see Figures\,\ref{fig:histograms} and\,\ref{fig:snr_comparison}). 

We have compared our non-iterative MMF6 catalogue with the official \textit{Planck} MMF3 catalogue, as both are obtained using the same frequency channels and the same standard MMF (albeit with several differences, such as the multipole range used and the use of a different \textit{Planck} data release). Both catalogues are broadly in agreement, featuring similar number counts both as a function of signal-to-noise and redshift, and similar signal-to-noise values for the common detections between the two (see Figure\,\ref{fig:comparison_szifi_mmf3}).


We have carefully assessed the impact of the cluster-correlated CIB on the retrieved cluster observables by comparing the Compton-$y$ amplitude $\hat{y}_0$ values that we have re-extracted with our 10 different MMFs at a fixed angular scale of 5\,arcmin for the clusters detected with all 10 MMFs (our CIB subsample). This comparison procedure was chosen in order to minimise selection effects. The differences in these fixed $\hat{y}_0$ measurements across the different MMFs are very similar to those observed for synthetic data in which the correlation between the tSZ and the CIB fields is artificially removed (see Figures\,\ref{fig:fixed_data} and\,\ref{fig:fixed_sim_random}). These findings have led us to conclude that the cluster-correlated CIB likely has a negligible impact on the cluster tSZ observables for the clusters in our catalogues. They also suggest that the impact of the cluster-correlated CIB on cosmological constraints obtained with our baseline iMM6 catalogue (as well as, perhaps, with the MMF3 catalogue) may be negligible. (See, e.g., \citet{Rotti2021Giants}, where the CIB was proposed as a possible explanation for the low SZ mass bias parameter $1-b$ required to reconcile the \textit{Planck} MMF3 number counts with the \textit{Planck} primary CMB power spectra.) This is, however, a possibility that ought to be investigated rigorously at the cosmological analysis level. We note that this assessment is the first of its kind for a \textit{Planck} cluster sample, not having been performed in the official \textit{Planck} analyses.

In addition, for the first time in the context of cluster finding, we have quantified the impact of the relativistic corrections to the tSZ SED. We have done so by re-extracting $\hat{y}_0$ for a cluster subsample (the rSZ subsample) assuming a relativistic tSZ SED, using electron temperature estimates obtained with a simulation-calibrated Compton-$y$--temperature scaling relation. We have found a significant $\sim 5$--$10$\,\% change on the $\hat{y}_0$ values but a negligible impact on the signal-to-noise values (see Figures\,\ref{fig:fixed_rsz_yt} and\,\ref{fig:fixed_rsz_qt}). These findings allow us to conclude that the tSZ relativistic corrections will be negligible in any cosmological analysis of our catalogues in which the signal-to-noise is used as the tSZ mass observable.

Finally, we have validated our \texttt{SZiFi}-based cluster detection pipeline by applying it to synthetic \textit{Planck}-like data for our 10 MMFs. Most notably, we have been able to produce a theoretical prediction for the number counts in the validation catalogues across both signal-to-noise and redshift that successfully describes them (see Figure\,\ref{fig:validation_counts} and\,\ref{fig:validation_counts_all}). This validation exercise, which, to the authors' knowledge, is the first of its kind for tSZ-detected clusters, constitutes  an important stepping stone towards using our catalogues in a cosmological number-count analysis.

We have not attempted to relate the signal-to-noise or $\hat{y}_0$ measurements in our catalogues to the cluster mass, something that we leave for further work. We also leave to further work the study of other potentially significant effects different from those considered here. These include, e.g., contamination from radio emission, which could be assessed in a similar way to how we quantified the impact of the cluster-correlated CIB, but focusing on the lower frequency channels. Finally, given their high purity (greater than 97\,\% for a signal-to-noise threshold of 5.5 for most catalogues) and the fact that we can successfully produce a theory prediction for our validation catalogues, we note that our \textit{Planck} \texttt{SZiFi} catalogues are well suited for a cosmological number-count analysis. We are currently working on one, which we intend to publish within the next year.

\section*{Acknowledgements}

The authors would like to thank Anthony Challinor, Erik Rosenberg, Boris Bolliet, Joseph J. Mohr, Sebastian Bocquet, Matthias Klein, Lindsey Bleem, and Nicholas Battaglia for useful discussions, and Elizabeth Lee for providing the cluster Compton-$y$--temperature scaling relation parameter values. The authors would also like to thanks the Aspen Center for Physics, where part of this work was completed.

ÍZ acknowledges support from the STFC (grant numbers ST/W000977/1 and ST/T000414/1). JC was supported by the ERC Consolidator Grant {\it CMBSPEC} (No.~725456) and by the Royal Society as a Royal Society University Research Fellow at the University of Manchester, UK (No.~URF/R/191023).

This work was performed using resources provided by the Cambridge Service for Data Driven Discovery (CSD3) oper- ated by the University of Cambridge Research Computing Service (\href{csd3.cam.ac.uk}{\texttt{csd3.cam.ac.uk}}), provided by Dell EMC and Intel using Tier-2 funding from the Engineering and Physical Sci- ences Research Council (capital grant EP/T022159/1), and DiRAC funding from the Science and Technology Facilities Council (\href{dirac.ac.uk}{\texttt{dirac.ac.uk}}), within the DiRAC Cosmos dp002 project.

In this project the authors have made use of the M2C Galaxy Cluster Database, constructed as part of the ERC project M2C (The Most Massive Clusters across cosmic time, ERC Advanced Grant No. 340519).

\section*{Data Availability}

The \textit{Planck} \texttt{SZiFi} master catalogue will become publicly available at \href{https://github.com/inigozubeldia/szifi/tree/main/planck_szifi_master_catalogue}{this link}. Before we make it available, it will be shared upon reasonable request. The synthetic data used throughout this work will also be shared upon reasonable request. The cluster finder used in this work, \texttt{SZiFi}, is already publicly available at \texttt{\href{https://github.com/inigozubeldia/szifi/}{github.com/inigozubeldia/szifi}}.



\bibliographystyle{mnras}
\bibliography{references} 




\appendix

\section{The \textit{Planck}  \texttt{SZiFi} master catalogue: catalogue entries}\label{appendix:catalogue}

\begin{table*}
\centering
\begin{tabular}{ll}
\hline

\textbf{Entry} & \textbf{Description} 
\\
\hline

\texttt{name} & Name of the detection. \\
\texttt{SNR\_} + \texttt{nameMMF} & Detection signal-to-noise $q_{\mathrm{obs}}$ for the MMF \texttt{nameMMF}, where \texttt{nameMMF} can be: \texttt{iMMF6}, \texttt{iMMF5}, \texttt{iMMF4}, \\ & \texttt{sciMMF6},  \texttt{sciMMF5}, \texttt{sciMMF4}, \texttt{sciMMF6\_beta}, \texttt{sciMMF6\_betaT}, \texttt{sciMMF5\_beta}, \texttt{sciMMF5\_betaT}. Add\\&  \texttt{noit} before \texttt{nameMMF} to  retrieve the non-iterative signal-to-noise.\\
\texttt{y0\_} + \texttt{nameMMF} & Detection $\hat{y}_0$. \\
\texttt{theta500\_} + \texttt{nameMMF} & Detection angular scale $\hat{\theta}_{500}$ in acrmin. \\
\texttt{tile\_} + \texttt{nameMMF} & Detection selection tile. \\
\texttt{glon\_} + \texttt{nameMMF} & Detection Galactic longitude. \\
\texttt{glat\_} + \texttt{nameMMF} & Detection Galactic latitude. \\
\texttt{thetax\_} + \texttt{nameMMF} & Detection $x$ coordinate in the local selection tile frame in rad. \\
\texttt{thetay\_} + \texttt{nameMMF} & Detection $y$ coordinate in the local selection tile frame in rad. \\
\texttt{SNR\_fixed\_} + \texttt{nameMMF} & SNR extracted at angular scale of 5\,arcmin and mean sky position (see Section\,\ref{sec:cib}).\\
\texttt{y0\_fixed\_} + \texttt{nameMMF} & $\hat{y}_0$ extracted at angular scale of 5\,arcmin and mean sky position.\\
\texttt{SNR\_fixed\_rel\_} + \texttt{nameMMF} & SNR extracted at angular scale of 5\,arcmin and mean sky position assuming the relativistic tSZ \\&  SED (see Section\,\ref{sec:rsz}).\\
\texttt{y0\_fixed\_rel\_} + \texttt{nameMMF} & $\hat{y}_0$ extracted at angular scale of 5\,arcmin and mean sky position assuming the relativistic tSZ SED.\\
\texttt{glon} & Galactic longitude, assigned following the master catalogue construction hierarchy.\\
\texttt{glat} & Galactic latitude, assigned following the master catalogue construction hierarchy.\\
\texttt{ra} & Right ascension, obtained from \texttt{glon} and \texttt{glat}.\\
\texttt{dec} & Declination, obtained from \texttt{glon} and \texttt{glat}.\\
\texttt{cosmology\_mask} & Whether the detection is contained within the cosmology mask (\texttt{1} if so, \texttt{-1} otherwise).\\
\texttt{cross-matched} & Whether cross-match has been found with the external catalogues (\texttt{1} if so, \texttt{-1} otherwise).\\
\texttt{confirmed} & Whether the detection is a confirmed cluster (\texttt{1}), an unconfirmed object (\texttt{0}), or a confirmed false \\ & detection (\texttt{-1}); see Section\,\ref{sec:confirmation}.\\
\texttt{flag\_gcc} & Whether a cross-match has  been found with a \textit{Planck} GCC (\texttt{1} if so, \texttt{-1} otherwise).\\
\texttt{redshift} & Cluster redshift. \\
\texttt{redshift\_type} & Redshift type (\texttt{spec}, \texttt{phot}, or \texttt{-1}).\\
\texttt{temperature} & Compton-$y$-weighted temperature estimate in keV (see Section\,\ref{sec:rsz}).\\ 
\texttt{name\_mcsz} & Name in the MCSZ meta-catalogue.\\
\texttt{name\_mcxc} & Name in the MCXC-II meta-catalogue.\\
\texttt{name\_actdr5} & Name in the ACT DR5 catalogue.\\
\texttt{name\_codex} & Name in the CODEX catalogue.\\
\texttt{name\_rass-mcmf} & Name in the RASS-MCMF catalogue.\\
\texttt{name\_erass} & Name in the eRASS catalogue.\\
\hline
\end{tabular}
\caption{Description of the entries in the \textit{Planck} \texttt{SZiFi} master catalogue.}
\label{table:catalogue}
\end{table*}

In Table\,\ref{table:catalogue} we offer a brief description of all the entries in the \textit{Planck} \texttt{SZiFi} master catalogue.

\section{Fixed Compton-$y$ results for a mass-selected cluster sample}\label{appendix:mass_selected}

In this appendix we offer further investigation on the origin of the two features seen in the re-extracted fixed $\hat{y}_0$ measurements for the randomised Websky catalogues that are discussed in Section\,\ref{sec:cib} (see, in particular, Figure\,\ref{fig:fixed_sim_random}). We recall that $\hat{y}_0$ is the MMF amplitude parameter, and that these features are: (i) a small shift in $\hat{y}_0$ for some MMFs for large $\hat{y}_0$ values and (ii) the shape of the $\hat{y}_0$ distribution at the low-$\hat{y}_0$ end. We do this by extracting $\hat{y}_0$ from our synthetic Websky-based CIB-randomised maps for all our 10 iterative MMFs for a mass-selected cluster sample, instead of the sample analysed in Section\,\ref{sec:cib}. In particular, we consider the Websky halo catalogue with $M_{500} > 6 \times 10^{14} M_{\odot}$, which contains 557 objects. We extract $\hat{y}_0$ using the \texttt{SZiFi}-based re-extraction pipeline presented in Section\,\ref{sec:cib} at the true halo sky positions and at a fixed angular size of $5$\,arcmin, the latter following what was done in Section\,\ref{sec:cib}.

\begin{figure}
\centering
\includegraphics[width=0.5\textwidth,trim={0mm 0mm 0mm 0mm},clip]{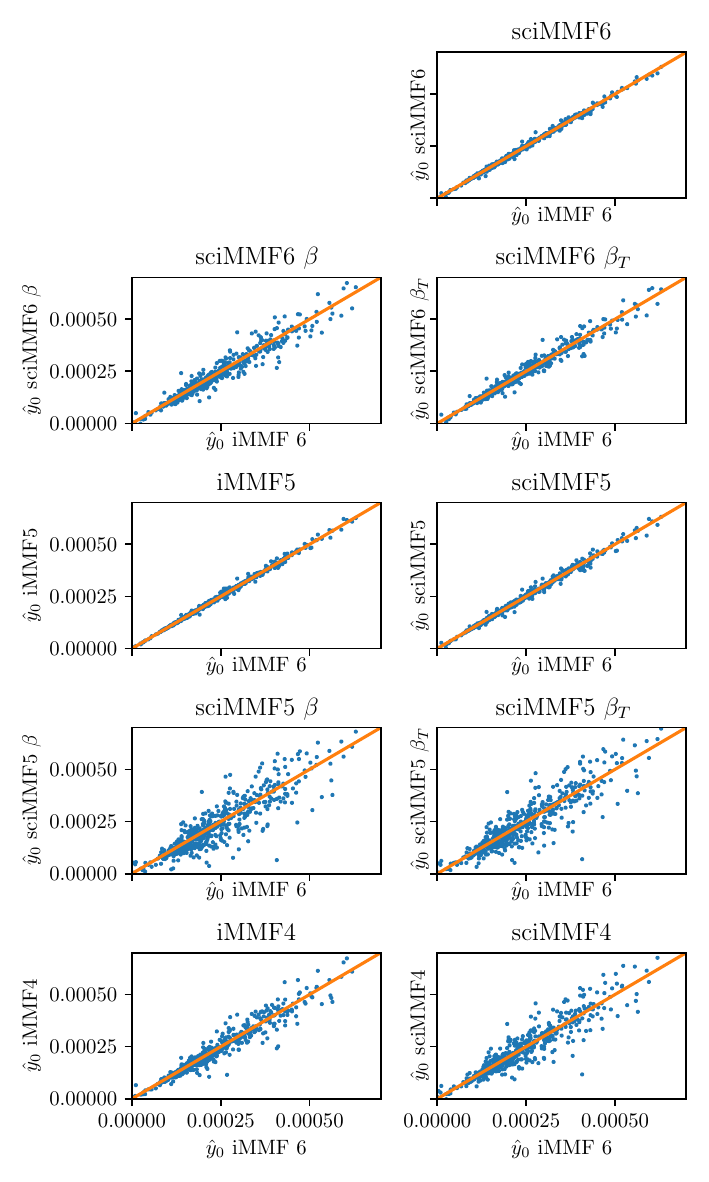}
\caption{Re-extracted fixed measurements of the MMF amplitude parameter $\hat{y}_0$ obtained from our \textit{Planck}-like Websky-based synthetic maps with the randomised CIB for a mass selected sample. In each panel, the $\hat{y}_0$ measurements for the panel's MMF are plotted against those for iMMF6 (vertical and horizontal axes, respectively). Excellent agreement across all MMFs is observed for the full $\hat{y}_0$ range. Compare with the analogous measurements for a tSZ-selected sample in Figure\,\ref{fig:fixed_sim_random}.}
\label{fig:fixed_y0_mass_selected}
\end{figure}

Figure\,\ref{fig:fixed_y0_mass_selected} shows our measurements. As in Figures\,\ref{fig:fixed_data}, \ref{fig:fixed_sim_random}, and\,\ref{fig:fixed_sim_correlated}, each pannel shows the $\hat{y}_0$ measurements for the panel's MMF on the vertical axis and those for iMMF6 on the horizontal axis. No evidence for neither of the features is seen here, with the measurements for each MMF being consistent with those for iMMF6 across the full $\hat{y}_0$ range. Therefore, we can conclusively attribute the two features to the selection of the sample analysed in Section\,\ref{sec:cib}.

\section{Validation of cluster detection pipeline for all MMFs}\label{appendix:validation}

\begin{figure*}
\centering
\includegraphics[width=1.0\textwidth,trim={0mm 0mm 0mm 0mm},clip]{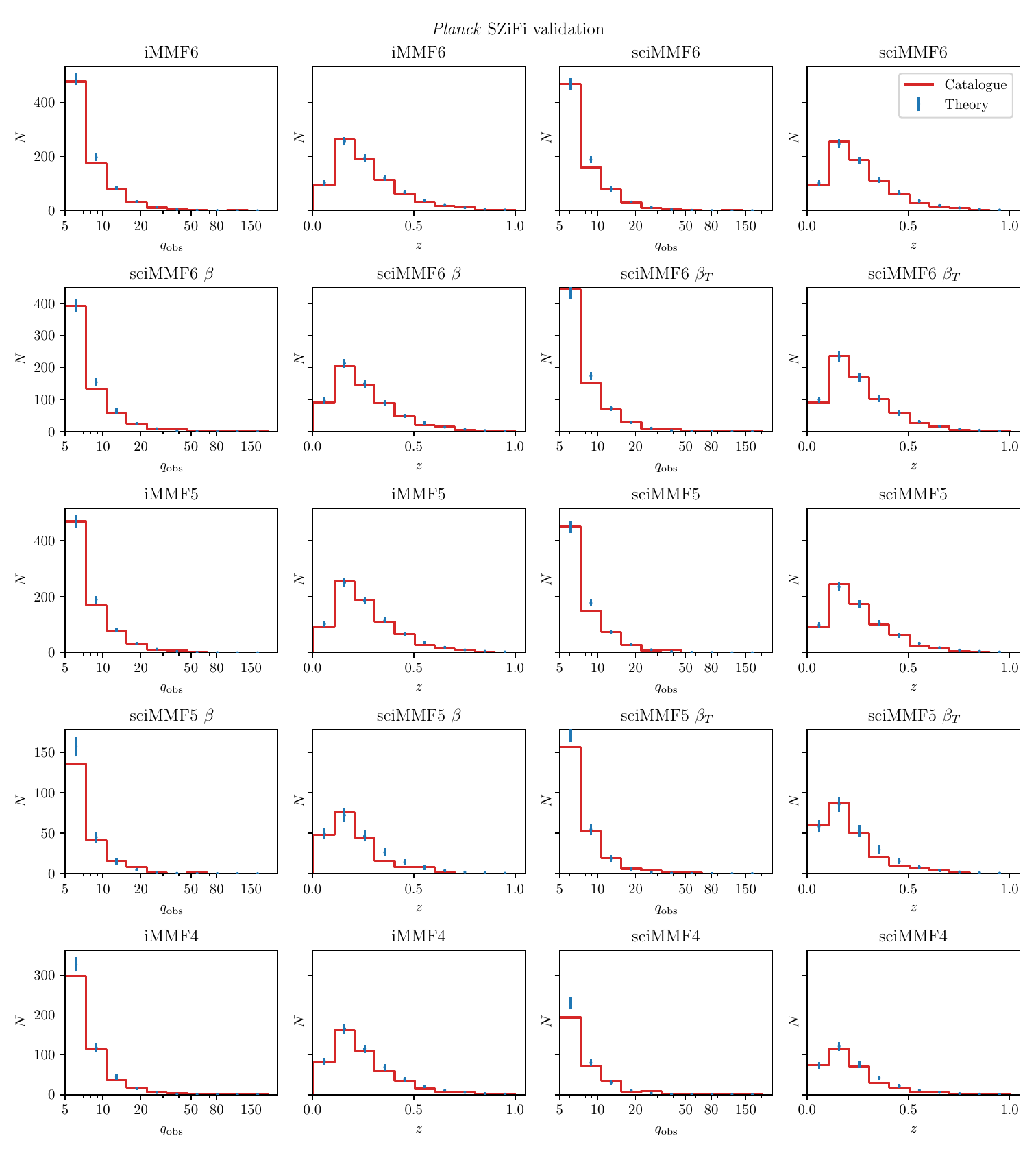}
\caption{Number counts a function of both signal-to-noise and redshift for all our iterative validation catalogues (red histograms), shown together with their corresponding theoretical prediction with the optimisation bias taken into account (blue data points). Only the confirmed (i.e., true) detections in the catalogues are considered.}
\label{fig:validation_counts_all}
\end{figure*}

Figure\,\ref{fig:validation_counts_all} shows the number counts across both signal-to-noise and redshift for the confirmed clusters in all our iterative validation catalogues (red histograms) together with the corresponding theoretical prediction with the optimisation bias taken into account (blue data points). As was the case for the iMMF6 validation catalogue (see Figure\,\ref{fig:validation_counts}), our theoretical model provides, overall, a good description of the observed counts.

\bsp	
\label{lastpage}
\end{document}